\documentclass{aa}
\usepackage[varg]{txfonts}

\usepackage{natbib,twoopt}
\usepackage[breaklinks=true]{hyperref} 
\bibpunct{(}{)}{;}{a}{}{,} 
\makeatletter
\newcommandtwoopt{\citeads}[3][][]{\href{http://adsabs.harvard.edu/abs/#3}%
{\def\hyper@linkstart##1##2{}%
\let\hyper@linkend\@empty\citealp[#1][#2]{#3}}}
\newcommandtwoopt{\citepads}[3][][]{\href{http://adsabs.harvard.edu/abs/#3}%
{\def\hyper@linkstart##1##2{}%
\let\hyper@linkend\@empty\citep[#1][#2]{#3}}}
\newcommandtwoopt{\citetads}[3][][]{\href{http://adsabs.harvard.edu/abs/#3}%
{\def\hyper@linkstart##1##2{}%
\let\hyper@linkend\@empty\citet[#1][#2]{#3}}}
\newcommandtwoopt{\citeyearads}[3][][]%
{\href{http://adsabs.harvard.edu/abs/#3}
{\def\hyper@linkstart##1##2{}%
\let\hyper@linkend\@empty\citeyear[#1][#2]{#3}}}
\makeatother

\usepackage{xcolor}
\usepackage{amsmath}

\begin{document}

\title{
 Idealised 3D simulations of diabatically forced Ledoux convection. Application to the
 atmosphere of hot rocky exoplanets
}




\author{
 Simon Daley-Yates\inst{1}
 \and
 Thomas Padioleau\inst{1}
 \and
 Pascal Tremblin\inst{1}
 \and
 Pierre Kestener\inst{1}
 \and
 Martial Mancip\inst{1}
}

\offprints{Simon Daley-Yates, \email{simon.daley@cea.fr}}

\institute{
 Universit\'e Paris-Saclay, UVSQ, CNRS, CEA, Maison de la Simulation, 91191,
 Gif-sur-Yvette, France
}


\date{Received ... / Accepted ...}

\abstract {}
{
 {We investigate the impact of dimensionality, resolution, and long
   timescales on convective numerical simulations forced by thermo-compositional diabatic processes. We focus our study on simulations that are stable to the Schwarzschild criterion but unstable to the Ledoux one (i.e. simulations with a stabilising temperature gradient and a destabilising mean-molecular-weight gradient). We aim to establish the possibility of a reduced temperature gradient in such setups.}
}
{
 A suite of 3D simulations incorporating both time series and convergence studies were conducted using a high-performance numerical hydrodynamic code. We used, {as a simplified and idealised test case}, a sample region of the secondary atmosphere of a hot rocky exoplanet, of the order of the scale height of the system, within which the chemical transition CO+O $\leftrightarrow$ CO$_{2}$ could occur. Newtonian cooling was employed to force an equilibrium temperature, and a chemical source term was used to maintain a negative mean-molecular-weight gradient in the vertical direction.
}
{
 Our results demonstrate that a mean-molecular-weight gradient and a chemical source term can reduce the atmosphere temperature gradient, a result that does not converge away with resolution and {is stable when exploring long timescales. Simulations in two dimensions are prone to the development of shear modes, as already seen in the literature for double-diffusive convection. The 3D convective steady state is not impacted by these shear modes, suggesting that this phenomenon is linked to the dimensionality of the problem. We also show that the presence of the reduced temperature gradient is a function of the forcing timescales, disappearing if the chemical forcing is too slow.} We find that the above transition leads to a bifurcation of the atmosphere's temperature profile {when the chemical forcing is fast. Such a bifurcation is reminiscent of the bifurcation seen in the boiling crisis for steam or liquid convection.}
}
{
 {With the reduced temperature gradient in these idealised setups,} there exists the possibility of an analogy of the reddening (currently observed in the spectra of brown dwarfs) in the spectra of rocky exoplanet atmospheres. This possibility needs, however, to be checked with detailed 1D models in order to precisely characterise the equilibrium thermal and compositional gradients, the thermal and compositional forcing timescales, and the impact of a realistic equation of state to, in turn, assess if the regime identified here will develop in realistic situations. However, the possibility of this reddening cannot be excluded a priori. This prediction is new for terrestrial atmospheres and represents strong motivation for the use of diabatic models when analysing the atmospheric spectra of rocky exoplanets that will be observed with, for example, {\it the James Webb Space Telescope}.
}

\keywords{
 planets and satellites: atmospheres -- planets and satellites: terrestrial planets
}

\titlerunning{Diabatic convection in hot rocky exoplanet atmospheres}

\maketitle

\section{Introduction}

Natural convection is a fundamental process governing behaviour at vastly differing length scales, from water boiling in a kettle to the interiors of stars \citep{Busse1978, Hurlburt1984, Bodenschatz2000}. Convection leads to the dissipation of energy, the mixing of fluids, and the transfer of heat, all vital processes in astrophysics but especially so in the context of stellar interiors \citep{Denissenkov2010, Brown2013, Garaud2015a} and (exo-)planetary and brown dwarf atmospheres \citep[e.g.][]{Saumon2008, Morley2012, Tremblin2015, Tremblin2017}.

Assessing the impact of thermal and compositional source terms (i.e. diabatic processes) has been a challenge, one that was recently tackled in \cite{Tremblin2019} as part of a general theory for thermo-compositional diabatic convection. In the above study, this theory was demonstrated to be applicable to thermohaline convection in Earth's oceans \citep{Stern1960, Turner1967, Gregg1988}, moist convection in Earth's atmosphere \citep{Arakawa2011, Yano2014}, and, potentially, $\rm{CO}$ and $\rm{CH}_{4}$ radiative convection in the atmospheres of giant exoplanets and brown dwarfs \citep{Tremblin2015, Tremblin2016, Tremblin2017, Tremblin2019}. The purpose of our study is to further establish the theoretical standing of thermo-compositional diabatic convection, proposed by \cite{Tremblin2019}, by investigating the impact of dimensionality, resolution, and long timescales on convective numerical  simulations forced by thermo-compositional diabatic processes. As an idealised test case, we chose a potential $\rm{CO+O}~\leftrightarrow~\rm{CO}_{2}$ transition in the secondary atmosphere of terrestrial exoplanets.

The composition and dynamics of rocky exoplanet atmosphere  are currently receiving considerable attention \citep{Ito2015,Madhusudhan2019, Modirrousta-Galian2021}. This is partially due to the imminent arrival of several next-generation telescopes, both space and ground based, which have the characterisation of exoplanet atmospheres in their remit. This is because of questions regarding the habitability and the imprint of the formation and evolutionary history of the planet \citep{Stevenson2016, Tinetti2018}.

The habitability of a planet's atmosphere depends directly on the separate stages of its evolution; in the cases of Earth and Mars,  for example, the early atmospheres were characterised by gaseous outflows from the forming planetary crust \cite{Elkins-Tanton2008}. In this type of environment there exists strong chemical gradients and high temperatures, due either to irradiation or to early evolutionary stages. This means that chemical reactions are abundant, for example the $\rm{CO+O}~\leftrightarrow~\rm{CO}_{2}$ transition. Such a transition is likely to have occurred early during the formation of the Earth's atmosphere as CO$_{2}$ and CO are abundant via outgassing \cite{Forget2014}. Understanding this mechanism is therefore not only interesting in the astrophysical context but also from the paleo-climatology prospective.

For hot secondary atmospheres of young or highly irradiated rocky exoplanets, we can assume that the vertical temperature gradient is negative (at least on the night side of tidally locked irradiated planets). This negative vertical temperature gradient, together with temperature-dependent chemical reactions, generally ensures that higher mean molecular weight species form above lower mean molecular weight species. In short, thermally driven chemistry often leads to heavier molecules above lighter molecules and, therefore, a positive vertical mean-molecular-weight gradient. In the context of the present study, this results in a CO to CO$_{2}$ transition from the deep  to the upper atmosphere. In this physical situation, the atmosphere is unstable to Ledoux convection (otherwise known as natural convection). The denser CO$_{2}$-dominated gas will fall and displace the lighter, less dense CO-dominated gas, which will in turn rise and replace the CO$_{2}$, resulting in an atmosphere inversion. This trivial situation would then remain, with heavier material below lighter material. However, as the mean molecular weight inversion proceeds, the chemical reaction mentioned above acts to replenish both the CO$_{2}$ at the top and the CO at the bottom of the atmosphere, restoring the unstable gradient. The two processes, inversion and chemical reaction, compete until an equilibrium is achieved and a continuous convective cycle is established.

Using this $\rm{CO+O}~\leftrightarrow~\rm{CO}_{2}$ transition as a test case, we have explored a range of atmosphere temperatures at low resolution in both 2D and 3D and one specific temperature at high resolution. We investigate both the susceptibility to convection at different CO-CO$_{2}$ mixtures and at different forcing timescales, as well as the system's response to small-scale flow features, which are accessible at high resolution. In these idealised setups, we show that a temperature-gradient reduction can emerge when the compositional forcing timescale is sufficiently fast, potentially leading to the equivalent of the reddening seen in brown dwarf atmospheres. For more realistic situations, further detailed 1D studies are needed to verify that the compositional and thermal gradients, forcing timescales, and a realistic equation of state (EOS) will indeed lead to such a reduced gradient. Despite these additional checks, the possibility cannot be excluded a priori and has observational consequences that could be tested with, for example, the James Webb Space Telescope (JWST).

In the following sections we first describe the numerical setup used to study the relevant convective regimes. We then describe and discuss the simulation results and its limitations, before reaching our conclusions.

\section{Numerical setup}

In the following sections we follow the analytic modelling proposed by \cite{Tremblin2019}, giving justification for the numerical regime and initial conditions that the simulations are conducted in.

\subsection{Governing equations}

The equations of hydrodynamics can be represented as the following hyperbolic partial differential equation:
\begin{equation}
 \partial_{t} \boldsymbol{U} + \boldsymbol{\nabla} \boldsymbol{F} =
 \boldsymbol{S}
 \label{eq:Euler}
,\end{equation}
where $\boldsymbol{U}$, $\boldsymbol{F}$, and $\boldsymbol{S}$ are the state vector of conserved variables, flux vector, and vector of source terms, respectively, and are defined as:
\begin{equation}
 \boldsymbol{U} = \begin{pmatrix} \rho \\ \rho \boldsymbol{u} \\ \rho X \\ \rho
  {\varepsilon}\end{pmatrix}, \ \boldsymbol{F} = \begin{pmatrix} \rho \\ \rho
  \boldsymbol{u} \otimes \boldsymbol{u} + P \\ \rho X \boldsymbol{u}
  \\ \boldsymbol{u} \left( \rho {\varepsilon} + P \right)\end{pmatrix},
 \ \boldsymbol{S} = \begin{pmatrix} 0  \\ \rho \pmb{\textsl{g}} \\ \rho R(X, P,
  T) \\ \rho c_{p} H(X, P, T)\end{pmatrix},
 \label{eq:Euler_seperate}
\end{equation}
where $\rho$, $\boldsymbol{u}$, $X$, $P$, $\varepsilon$, $c_{p}$, and $T$ are the density, velocity, mass-mixing ratio, pressure, total energy, and specific heat capacity at constant pressure and temperature, respectively. This system is closed using the ideal gas law,
\begin{equation}
 P = \frac{\rho k_{\rm{B}} T}{\mu(X) m_{\rm{H}}},
 \label{eq:state}
\end{equation}
where $m_{\rm{H}}$ is the mass of Hydrogen. The total energy is $\varepsilon = e + u^{2}/2 + \phi$, where $e$ is the internal energy and $\phi$ the gravitational potential, with $\pmb{\textsl{g}} = - \boldsymbol{\nabla} \phi$ the gravitational acceleration vector. The mean molecular weight is related to the mass-mixing ratio via
\begin{equation}
 \frac{1}{\mu(X)} = \frac{X}{\mu_{1}} + \frac{1 - X}{\mu_{2}},
 \label{eq:mean_mol}
\end{equation}
where $\mu_{1}$ and $\mu_{2}$ are the mean molecular weights of the two fluid species. {We assume in this numerical study a constant adiabatic index $\gamma$ in order to finely control the convective regime of the system and ensure that the simulations will be Schwarzschild stable and Ledoux unstable (i.e. will have a stabilising temperature gradient and de-stabilising mean-molecular-weight gradient). The purpose of this choice is to assess the impact of the mean-molecular-weight gradient alone in the simulations, we discuss in Sect.~\ref{sect:result} the limitations of this choice for realistic test cases. With a constant adiabatic index, the EOS can be rewritten as $P=\rho e(\gamma-1)$, with $e=c_v T$, and $c_v$ the specific heat capacity at constant volume: $c_v=(k_{\rm{B}}/\mu(X) m_{\rm{H}}(\gamma-1))$.}

The $\rho R(X, P, T)$ is a source term for the equation governing the advection of the mass-mixing ratio. Physically it represents the conversion between chemical species. $\rho c_{p} H(X, P, T)$ is the source term for the energy equation and represents energy input into the system via numerous sources, for example, radiation, thermal diffusion or release of latent heat via chemical reactions. If both $R(X, P, T)$ and $H(X, P, T)$ are non-zero, then the system in general is non-conservative.

\subsection{Diabatic and adiabatic convection}

For a fluid obeying an ideal gas law and under the influence of an external plane parallel gravitational field and experiencing no other momentum, energy or mass source terms, the criterion for convection is:
\begin{equation}
 \nabla_{T} - \nabla_{\rm{ad}} > 0
 \label{eq:schwarzschild}
,\end{equation}
where $\nabla_{T}~=~-h_{p} \partial T_{0} / \partial z$ and $\nabla_{\rm{ad}}~=~(\gamma - 1)/\gamma$, with $1/h_{p}~=~-\partial P_{0} / \partial z$, $P_{0}$ is the initial vertical pressure profile, and $T_{0}$ is the initial temperature profile. This is known as Schwarzschild convection. In this regime, $\nabla_{\rm{ad}}~=~\rm{constant}$ (assuming an ideal gas law), a system's susceptibility to convection is determined solely by $\nabla_{T}$, the vertical temperature gradient. Therefore, we are free to set the system to be Schwarzschild stable simply by setting a temperature gradient such that $\nabla_{T} < \nabla_{\rm{ad}}$ at $t = t_{0}$. Such a system is then stable to convection for all $t > t_{0}$.

However, if there are multiple molecular species present in the fluid, the mean molecular weight need not be homogeneous. A gradient can be defined that represents a transition from one chemical species to another; $\nabla_{\mu}$. Modifying Eq. \ref{eq:schwarzschild} to account for $\nabla_{\mu}$ leads to the following expression:
\begin{equation}
 \nabla_{T} - \nabla_{\rm{ad}} - \nabla_{\mu} > 0.
 \label{eq:ledoux}
\end{equation}
This means that even if the Schwarzschild criterion for stability is met, the system can still be unstable. Equation \ref{eq:ledoux} is known as the Ledoux convection criterion. This criterion is referred as `adiabatic' convection in \cite{Tremblin2019} by opposition to the `diabatic' criterion that involves the chemical and thermal source terms. (See the appendix in \cite{Tremblin2019} for a full derivation of the instability criteria.) As proposed in \cite{Tremblin2019}, the conserved quantities in the presence of source terms are not potential temperature and composition but linear (logarithmic) combinations of both, similar to moist potential temperature in Earth atmosphere. {However, the situation is simpler in the context of Ledoux convection: the conserved quantity in the linear regime is simply proportional to the entropy. At saturation, the Ledoux criterion is equivalent to homogeneous entropy, hence the saturation is a nearly isentropic state. However, the chemical source term can still impact Ledoux convection in the non-linear regime by replenishing the mean-molecular-weight gradient. In that context, we can expect a reduction of the temperature gradient for a Ledoux unstable system susceptible to a chemical reaction with a nearly constant entropy profile (i.e. adiabatic profile) as shown with the 2D simulations presented in \cite{Tremblin2019}. We emphasise that the temperature-gradient reduction in the Ledoux regime (with the adiabatic criterion) is then different from the temperature-gradient reduction with the diabatic criterion (moist or other source terms) since it does not happen in a sub-adiabatic regime here.}

\subsection{Physical system}

\begin{table*}
 \caption{Simulation parameters.}
 \label{tab:params}
 \centering
 \begin{tabular}{c c c}
  \hline\hline
  Variable                       & Symbol                            & Value            \\
  \hline
  Deep-atmosphere temperature    & $T_{0}$                           & 2300 - 2700 K    \\
  Upper-atmosphere temperature   & $T_{1}$                           & 2100 - 2500 K    \\
  Base pressure                  & $P_{0}$                           & 1 Bar            \\
  Gravitational acceleration     & $\textsl{g}_{z}$                  & 1000 cm s$^{-2}$ \\
  Thermal timescale             & $\tau_{\rm{rad}}$                 & 100 s            \\
  Chemical timescale            & $\tau_{\rm{chem}}$                & 10 s             \\
  CO$_{2}$ mean molecular weight & $\mu_{1}$                         & 44 g mol$^{-1}$  \\
  CO+O mean molecular weight     & $\mu_{2}$                         & 22 g mol$^{-1}$  \\
  {Adiabatic index}          & $\gamma$                          & 5/3            \\
  Simulation volume              & $L_{x} \times L_{y} \times L_{z}$ & 20 km$^{3}$      \\
  Initial perturbation           & $A$                               & 10 cm s$^{-1}$   \\
  Total simulation time          & $t_{\rm{max}}$                    & $10^{4}$ s       \\
  \hline
 \end{tabular}
\end{table*}

As discussed in the introduction, the atmospheric chemical transition $\rm{CO+O}~\leftrightarrow~\rm{CO}_{2}$ is a representative process for exploring the role that diabatic convection plays in shaping thermodynamic structures in rocky exoplanet atmospheres. This chemical transition is consistent with the atmosphere of the early Earth and young or highly irradiated terrestrial exoplanets in general. As such, the magnitudes of the physical quantities used in this study are consistent with this context and have been estimated with the 1D atmospheric code \texttt{ATMO} \citep{Tremblin2015}. However, we emphasise that \texttt{ATMO} is not yet ready to perform precise models of secondary atmospheres due to the lack of the relevant collisionally induced absorptions. We therefore simply estimated the pressure and temperature ranges and timescales of the $\rm{CO+O}~\leftrightarrow~\rm{CO}_{2}$ in a high metallicity model ([M/H]$\sim$3). As a consequence, the simulations presented in this paper should be seen as an idealised numerical experiment {and we discuss further these limitations in Sect.~\ref{sect:result}.}

\subsubsection{Initial conditions}

\begin{figure}
 \centering
 \includegraphics[width=\linewidth]{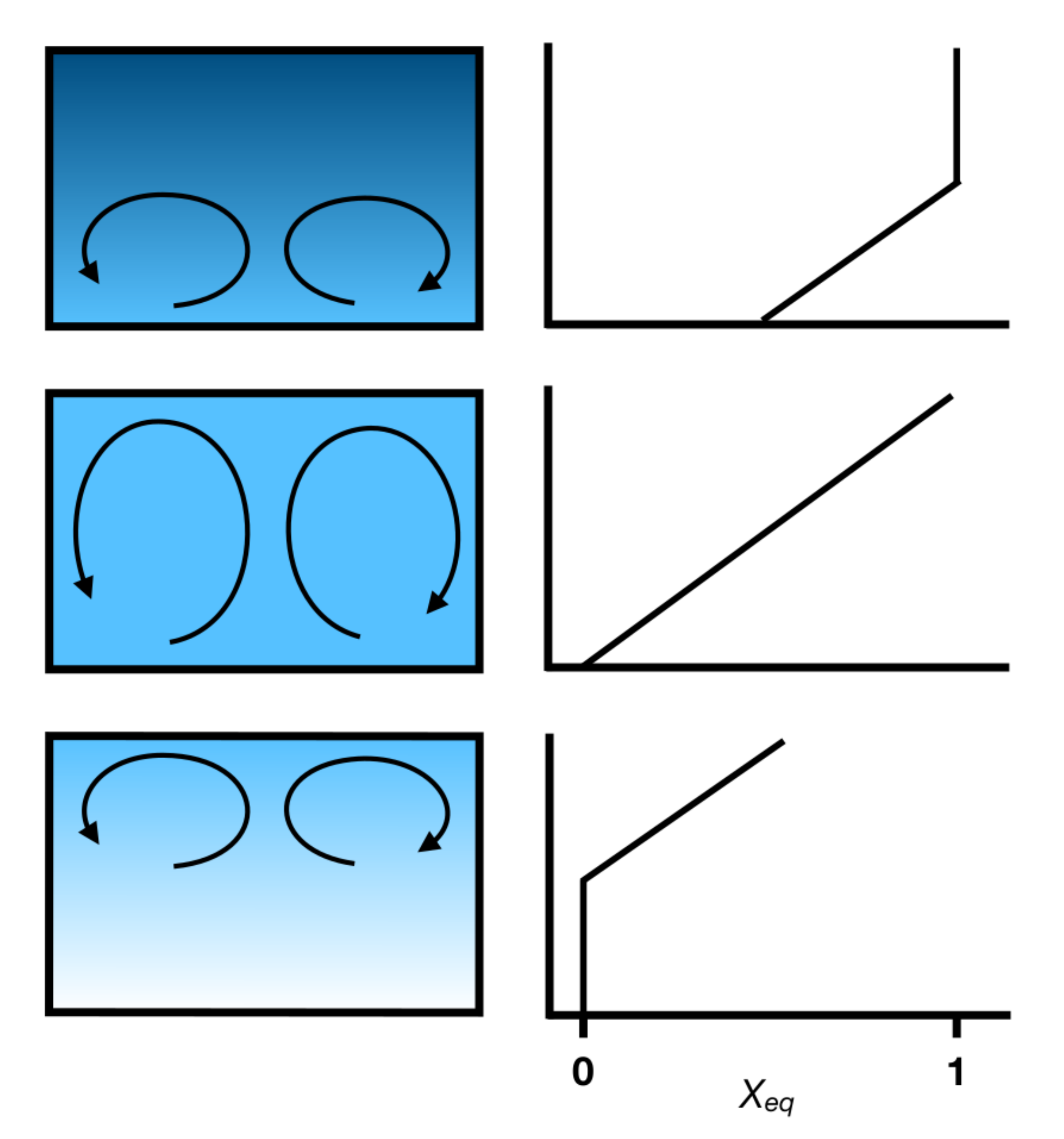}
 \caption{Diagram illustrating the parametrisation of the mass-mixing ratio and the subsequent impact on the convectively unstable region of the simulation domain. This parametrisation directly leads to a mean-molecular-weight gradient that drives the convection. At the limits of the temperature range explored, there is convective stability in the domain, as $\nabla \mu(X_{\rm{eq}}) = 0$. \label{fig:diagram}}
\end{figure}

The vertical temperature profile, $T(z)$, is specified \emph{ab initio} as a linear transition between pre-specified deep ($T_{0}$) and upper ($T_{1}$) atmosphere temperatures; thus,
\begin{equation}
 T_{\rm{eq}}(z) = T_{0} + z (T_{1} - T_{0}),
 \label{ref:temp_prof}
\end{equation}
resulting in a constant negative temperature gradient in the $z$ direction ($T_{0} > T_{1}$). A temperature gradient of $\Delta T~=~10^{-4} \ \rm{K} \ \rm{cm}^{-1}$ was kept the same for every simulated parameter set, only the absolute values were changed. In the following, the subscript $_{\rm{mid}}$ refers to the middle of the temperature range. So for the deep-atmosphere temperature, $T_{0, \rm{mid}} = 2500 \rm{K}$ and upper-atmosphere temperature, $T_{1, \rm{mid}} = 2300 \rm{K}$. In combination with the initial temperature profile, $T_{0}(z)$, we set the mass-mixing ratio profile as a parametrised transition from zero to unity while $T_{0, \rm{mid}}~<~T_{\rm{eq}}(z)~<~T_{1, \rm{mid}}$, thus:
\begin{equation}
 X_{\rm{eq}} \left( T_{\rm{eq}} \right) =
 \begin{cases}
  1 & \text{if } T_{\rm{eq}}(z) < T_{1, \rm{mid}} \\
  0 & \text{if } T_{\rm{eq}}(z) > T_{0, \rm{mid}} \\
  \frac{T_{\rm{eq}}(z) - T_{0, \rm{mid}} }{T_{1, \rm{mid}} - T_{0, \rm{mid}} }
    & \text{else}
 \end{cases}
.\end{equation}
If $T_{\rm{eq}}(z)$ is below (above) this range it is held at zero (unity). The result of this parametrisation is to impose a transition region in the mean molecular weight, which is calculated directly from the mass-mixing ratio, which in turn comes form the temperature profile. As we explore the deep-atmosphere temperature range, $T_{0} = 2300 K \rightarrow 2700 K$, the position of this transition region transits across the simulation domain from bottom to top. While this procedure may seam convoluted, the outcome is that at the extremities of the temperature range (2300 K or 2700 K) the mass-mixing ratio is either zero or unity everywhere within the simulation domain. This results in $\nabla \mu(X_{\rm{eq}}) = 0$ and hence convective stability (if $\nabla_{T}$ is such that Eq. (\ref{eq:ledoux}) $<0$). Fig. \ref{fig:diagram} illustrates this parametrisation and lack of convection in the regions where $X_{\rm{eq}} = \rm{constant}$.

All quantities, including $X_{\rm{eq}}$, are homogeneous in the $x$-$y$ plane. Hence, in the following, we have included the $k$ ($z$ direction) index but omitted the horizontal plane indices $j$ ($y$ direction) and $i$ ($x$ direction). Vertical hydrostatic equilibrium is achieved using a combination of the EOS (Eq. (\ref{eq:state})) and a finite difference approximation to the hydrostatic equation:
\begin{equation}
 \frac{P_{k} - P_{k-1}}{\Delta z} = -\frac{\rho_{k} + \rho_{k-1}}{2}
 g_z.
\end{equation}
Rearranging this expression and substituting Eq. (\ref{eq:state}), leads to the following pressure profile,
\begin{equation}
 p_{k} = p_{k-1} \frac{1 - \eta \frac{\mu(X)_{k-1}}{T_{k-1}}}{1 + \eta
  \frac{\mu(X)_{k}}{T_{k}}},
 \label{eq:prs_equil}
\end{equation}
where $\eta$ is a constant:
\begin{equation}
 \eta = \frac{\Delta z}{2} \frac{m_{\rm{H}}\textsl{g}_{z}}{k_{\rm{B}}}.
\end{equation}
The $\rho_{k}$ is then derived from Eq. (\ref{eq:state}).

Even if $\nabla \mu(X_{\rm{eq}}) \neq 0$, hydrostatic equilibrium means that an initial perturbation is required to seed convective motion. This is prescribed by:
\begin{align}
 \Delta \boldsymbol{u} & = A
 \begin{pmatrix}
  \sin(2 \pi \theta_{x})\cos(2 \pi \theta_{y})\sin(\pi \theta_{z}) \\
  \cos(2 \pi \theta_{x})\sin(2 \pi \theta_{y})\sin(\pi \theta_{z}) \\
  \cos(2 \pi \theta_{x})\cos(2 \pi \theta_{y})\cos(\pi \theta_{z})
 \end{pmatrix},
\end{align}
with $\theta_{x} = \frac{x - x_{\rm{mid}}}{L_{x}}$, $\theta_{y} = \frac{y - y_{\rm{mid}}}{L_{y}}$ and $\theta_{x} = \frac{z - z_{\rm{mid}}}{L_{z}}$ being the distances from the box midpoint in units of box width. This sets up a perturbation of two convective roles in the $x$ and $y$ directions, both spanning the complete height of the box. This is illustrated in the column of Fig. \ref{fig:diagram}

The form of Eq. \ref{eq:prs_equil} requires the specification of the deep-atmosphere temperature and pressure ($P_{0}$), the initial perturbation also requires an amplitude, $A$. These parameters are summarised in Table \ref{tab:params}.

\subsubsection{Thermal and chemical source terms}

Heating in the form of radiation transfer from the base of the atmosphere is mimicked by applying a Newtonian source term to the energy equation. This effectively forces the system towards an equilibrium state, $T_{\rm{eq}}$ (in this case the initial profile) on a timescale set by $\tau_{\rm{rad}}$ via the equation:
\begin{equation}
 \Delta T = \frac{T^{n} + T_{\rm{eq}} \frac{\Delta t}{\tau_{\rm{rad}}} }{1 +
  \frac{\Delta t}{\tau_{\rm{rad}}}} - T^{n}.
 \label{eq:delta_t}
\end{equation}
This change in temperature is then converted into a change in energy via the EOS and incorporated into the source term in Eq. \ref{eq:Euler_seperate}, $H(X, P, T)$.

In an identical manner, the mass-mixing ratio, $X$, is forced to equilibrium, $X_{\rm{eq}}$, according to:
\begin{equation}
 \Delta X = \frac{X^{n} + X_{\rm{eq}}(T^n) \frac{\Delta
   t}{\tau_{\rm{chem}}} }{1 + \frac{\Delta t}{\tau_{\rm{chem}}}} - X^{n},
 \label{eq:delta_x}
\end{equation}
on the timescale $\tau_{\rm{chem}}$. This expression is equivalent to $R(X, P, T)$. $T^{n}$ and $X^{n}$ are the states at the current time step, before the forcing is applied. The forced mass-mixing ratio is then used to compute the updated mean molecular weight. This process acts to maintain the spacial distribution of CO and CO$_{2}$ and is in competition with the advection of $X$. if no forcing terms were present, then mixing and diffusion would result in a homogeneous value of $X$ across the simulation domain. The role of Eq. (\ref{eq:delta_x}) is then to represent the capture, advection and subsequent release of energy through composition and the EOS in the system.

Both the heating of the system and the action of chemical reactions are constrained to occur on specific timescales, $\tau_{\rm{rad}}$ and $\tau_{\rm{chem}}$, respectively. These are parameterised according to the results of 1D calculations conducted using the ATMO code \citep{Tremblin2015} and have the values $\tau_{\rm{rad}}~=~100 \ \rm{s}$ and $\tau_{\rm{chem}}~=~10 \ \rm{s}$ (these values are also in Table \ref{tab:params}). {The estimations of these timescales follow the same procedure used in \cite{Tremblin2015} to estimate the fingering instability criterion.\ However, we should point out here that the chemical network used in ATMO is targeted for hydrogen-dominated planets \citep{Venot2012}.  These timescales are therefore rough estimates, and more detailed 1D modelling is needed to assess them properly for secondary atmospheres.}

{The radiative timescale, $\tau_{\rm{rad}}$, can be evaluated in a diffusion approximation as done in \cite{Tremblin2015}: it typically varies from orders of magnitude in the atmosphere from $10^{7}$ s at 100 bars down to 0.1 s at 0.01 bar. Close to 1 bar $\tau_{\rm{rad}}$ is of the order of 100 s. Using the ATMO code coupled to the out-of-equilibrium network of \cite{Venot2012}, we explored the chemical timescale, $\tau_{\rm{chem}}$, by adjusting $K_{\rm{zz}}$ to get a quenching point at 1 bar. The timescale deduced $K_{\rm{zz}}$ is of the order of 10 s but we point out that it can also vary by orders of magnitude in the atmosphere. since we use a chemical network and not a parametrisation we do not have easily access to the reaction timescales. As an indication, one can use the timescale for CO2 in the case of hydrogen-dominated planets that was computed in \cite{Zahnle2014}; it varies from $10^{-8}$ s at 100 bars to $10^{7}$ s at 0.01 bars for the pressure-temperature profile modelled with ATMO. These numbers cannot be used directly for our setup since we do not simulate a hydrogen-dominated planet but it gives an idea of how strongly the chemical timescale can vary. Based on the fact that both the chemical and radiative timescales vary strongly in the atmosphere and will be dependent on many details of the planet under consideration, we explore different cases in the paper, with a chemical timescale faster than the radiative timescale (base model) and vice versa.}

To determine the impact of chemical timescale, $\tau_{\rm{chem}}$, on the convergence of the simulation results; several simulations of the central deep-atmosphere temperature $T_{0} = 2522 \rm{K}$ were conducted. Four simulations were run with $\tau_{\rm{chem}}~=~1,~10,~100,~1000 \ \rm{s}$, the results of which are discussed in Sect. \ref{sec:FT}. However, the simulation suite that covered the full $T_{0}$ range was limited to $\tau_{\rm{chem}}~=~10$.

\subsubsection{Boundary conditions}

The physical extent of the simulation domain was bounded by several types of
conditions. Periodic boundary condition were imposed in the two horizontal
directions, $x$ and $y$. In the case of the vertical $z$ direction, a custom
condition was used for the lower and upper boundaries. At the upper and lower
atmosphere's boundaries, both the temperature and the passive scalar, $X$, was
allowed to float, meaning that both quantities in the boundary were extrapolated
from the computationally active region via a backwards or forwards (lower and
upper boundary, respectively) finite difference approximation, assuming a
constant gradient. {We use Neumann boundary conditions for the transverse
  velocities, and reflexive boundary conditions for the normal velocity at the
  lower and upper boundary.} Then the conservative variables are computed from
the temperature and mean molecular weight, $\mu$, (from the mass-mixing ratio)
and Eq. (\ref{eq:prs_equil}) and the ideal gas law. This results in
hydrostatic equilibrium enforced both below and above the simulation
domain. This setup results in a simulation domain where there is stable
non-convective atmosphere above and below the simulated region. {These
  boundary conditions ensure that the normal velocity at the interface of the
  domain are exactly zero, hence the simulations conserve the total mass at
  machine precision.}

\begin{figure}
 \centering
 \includegraphics[width=0.5\textwidth]{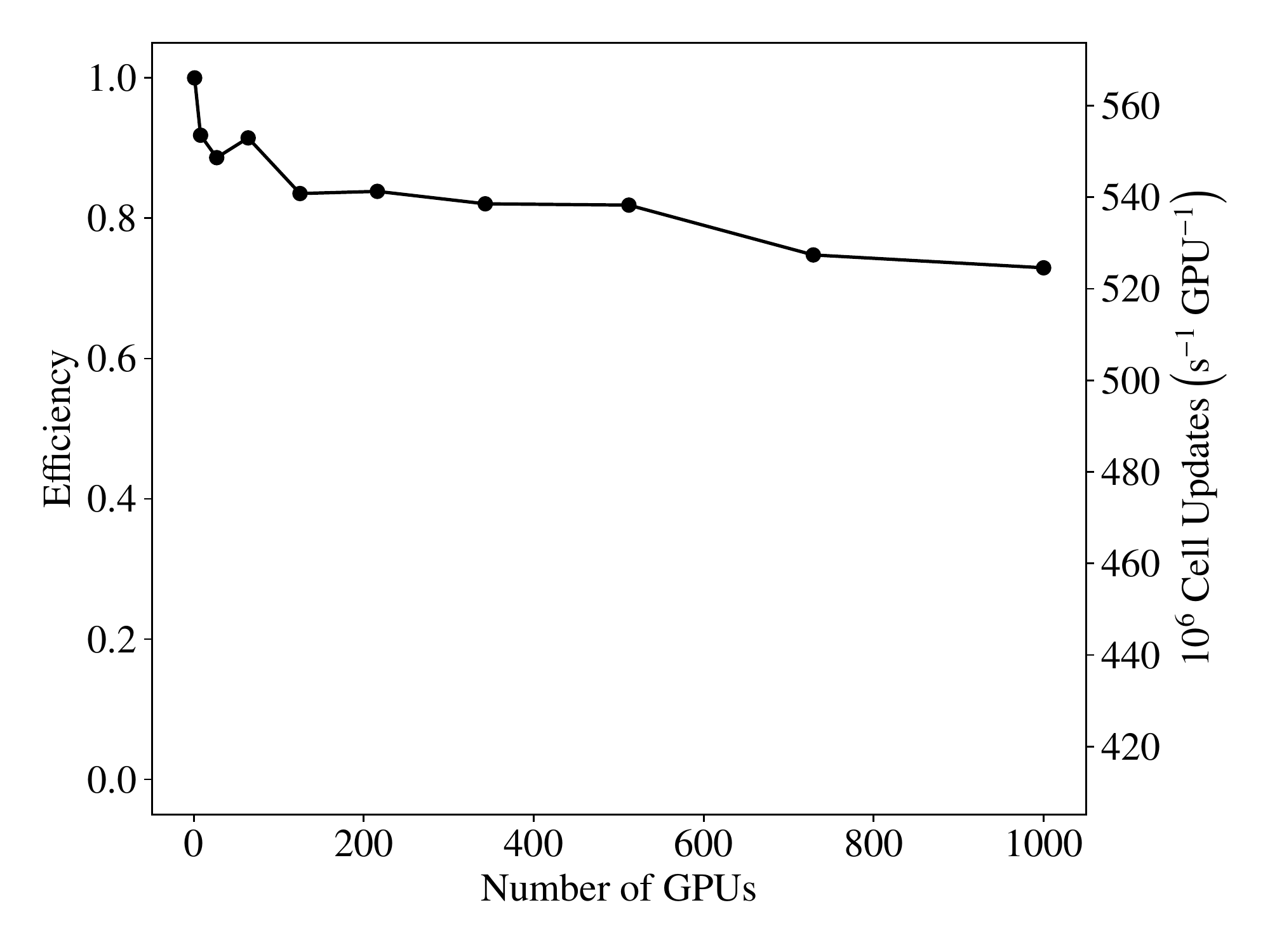}
 \caption{Weak scaling of the ARK code. Good efficiency is maintained for all numbers of GPGPUs, dropping to slightly lower than $80\%$ for 600 or greater GPGPUs. This scaling test was conducted with a numerical grid of 496$^{3}$ cells and used the same convection simulation as described in the text, but with a maximum iteration count of 1000. \label{fig:scaling}}
\end{figure}

\begin{figure*}
 \centering
 \includegraphics[width=1.0\textwidth]{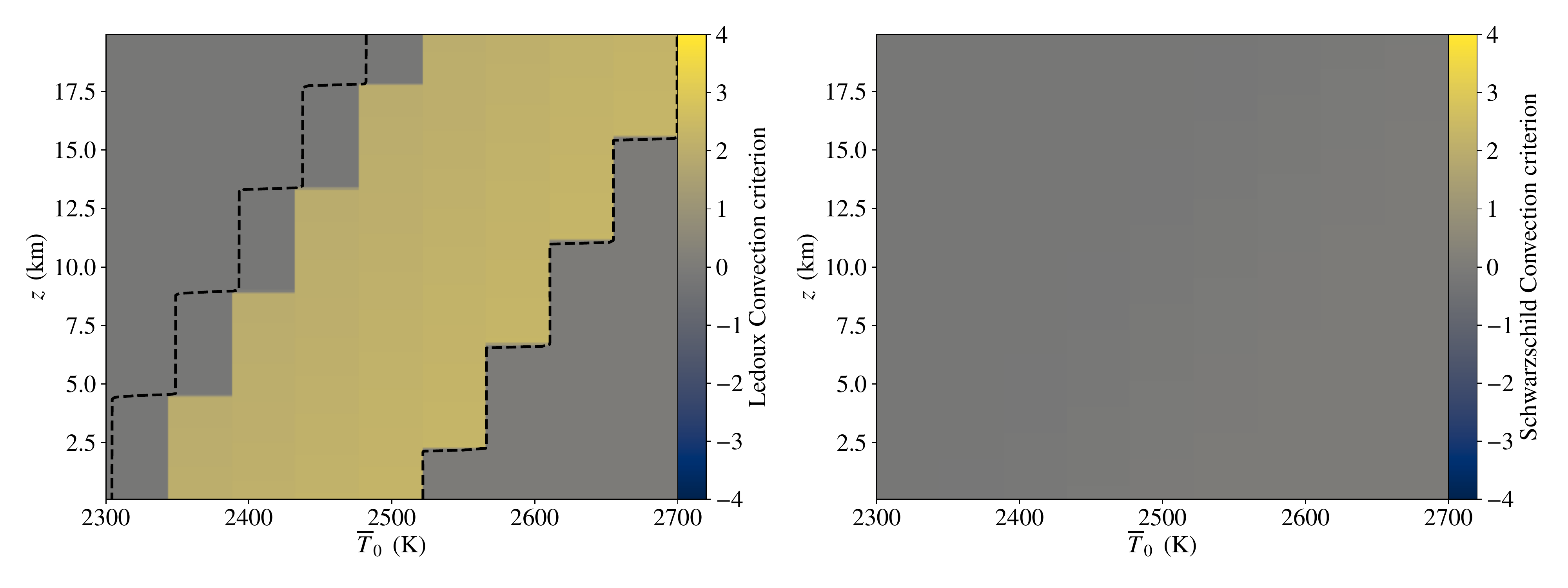}
 \includegraphics[width=1.0\textwidth]{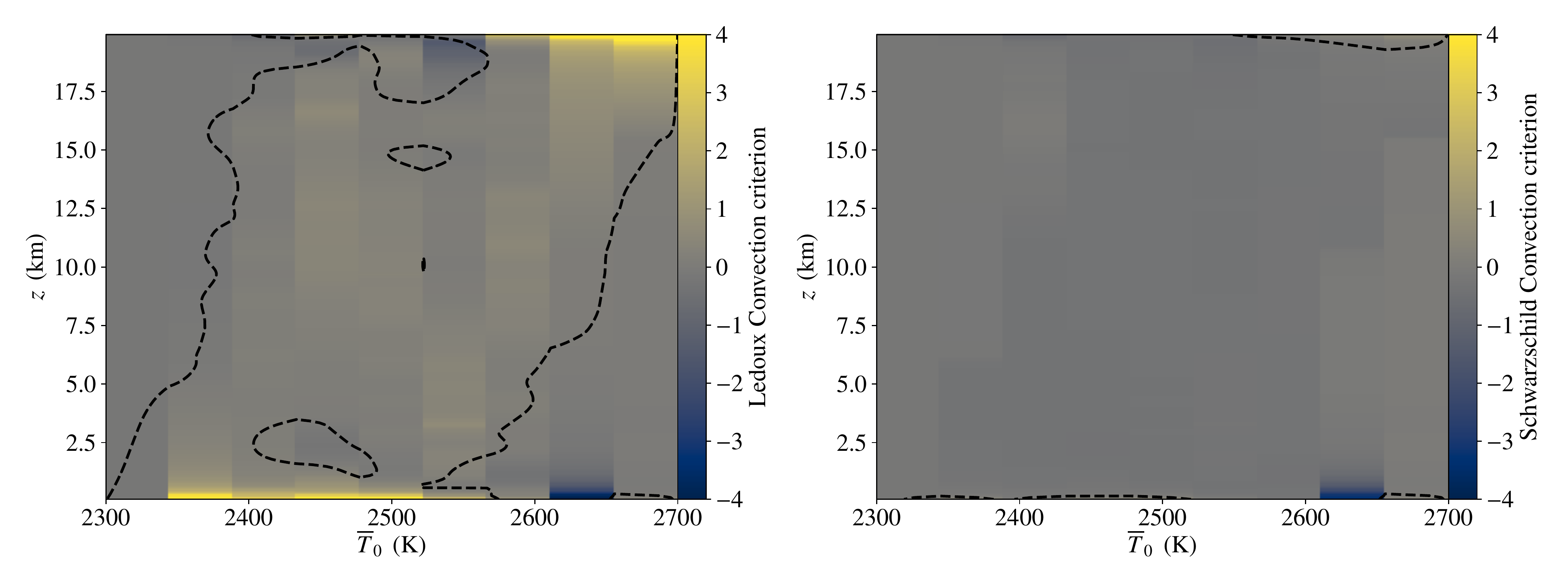}
 \caption{Pseudo-2D profiles constructed from 1D profiles of the criterion for Ledoux (left) and Schwarzschild (right) convection as a function of both deep-atmosphere temperature and altitude. The profiles are averaged in both the $x$ and $y$ directions. Every deep-atmosphere temperature value represents a different simulation. The dark blue contour indicates unity and hence the threshold for the convective instability. Where the criterion is greater than unity, the system is susceptible to convection. We can see from the position of this contour that the system is unstable to Ledoux convection for all $T_{0}$, both in the initial conditions and in steady state. This is not the case for Schwarzschild convection, with the initial conditions stable for all $T_{0}$, both initially and once the system is in equilibrium. \label{fig:criterion}}
\end{figure*}

\subsection{Simulations}

The numerical simulations were conducted using the ARK\footnote{\href{https://gitlab.erc-atmo.eu/erc-atmo/ark}{https://gitlab.erc-atmo.eu/erc-atmo/ark}} (all-regime Kokkos) code \citep{Padioleau2019}, which implements the all-regime method of \cite{Chalons2016a}. This approach allows for both low and high Mach number flows to be captured in the same simulation. A secondary feature of the code is the ability to capture hydrostatic balance and ensures it to machine precision. This property is why the method is described as being well-balanced. The codes ability to reproduce convection in both the Rayleigh-Benard and Schwarzschild regimes has been demonstrated by \cite{Padioleau2019}, making the code particularly well suited to exploring convection in the Ledoux regime.

ARK leverages the Kokkos framework for shared memory parallelism \cite{Edwards2014} in the C++ programming language. In essence, Kokkos allows for code to be written in a generic manner but be executed on multiple hardware architectures, while maintaining high numerical performance. This programming paradigm is known as performance portability, a property critical for the present study as it allows for the use of accelerators such as Intel KNLs and GPGPUs. Convergence of the simulations, at the resolution employed in a reasonable time frame, could only be achieved via the use of such accelerators.

\subsubsection{Computational setup}
\label{sec:num_setup}

The simulations are broken down into two distinct categories. The first are relatively low resolution (relative to the maximum resolution used) and are designed specifically to not be numerically intensive, allowing for a parameter study in temperature to be conducted. The second category concentrates exclusively on a single simulation, allowing for higher resolution, and thus serves as a convergence study to verify the validity of the low resolution simulations. In addition, this high resolution simulation acts to test the scalability of ARK on large numbers of GPGPUs.

This second category of simulation was carried out on the Jean Zay supercomputer of IDRIS under a GENCI allocation for `grand challenge'. This machine allowed us to run ARK on up to 1000 Nvidia V100 GPGPUs. A weak scaling study to test ARK on this architecture is presented in Fig. \ref{fig:scaling} and illustrates the efficiency of ARK at the scales afforded by Jean Zay. This scaling study was conducted with 496$^{3}$ numerical cells per GPGPU with each run lasting 1000 time steps.

\begin{figure}
 \centering
 \includegraphics[trim={0.5cm 2cm 0.5cm 0.5cm},clip,width=0.45\textwidth]{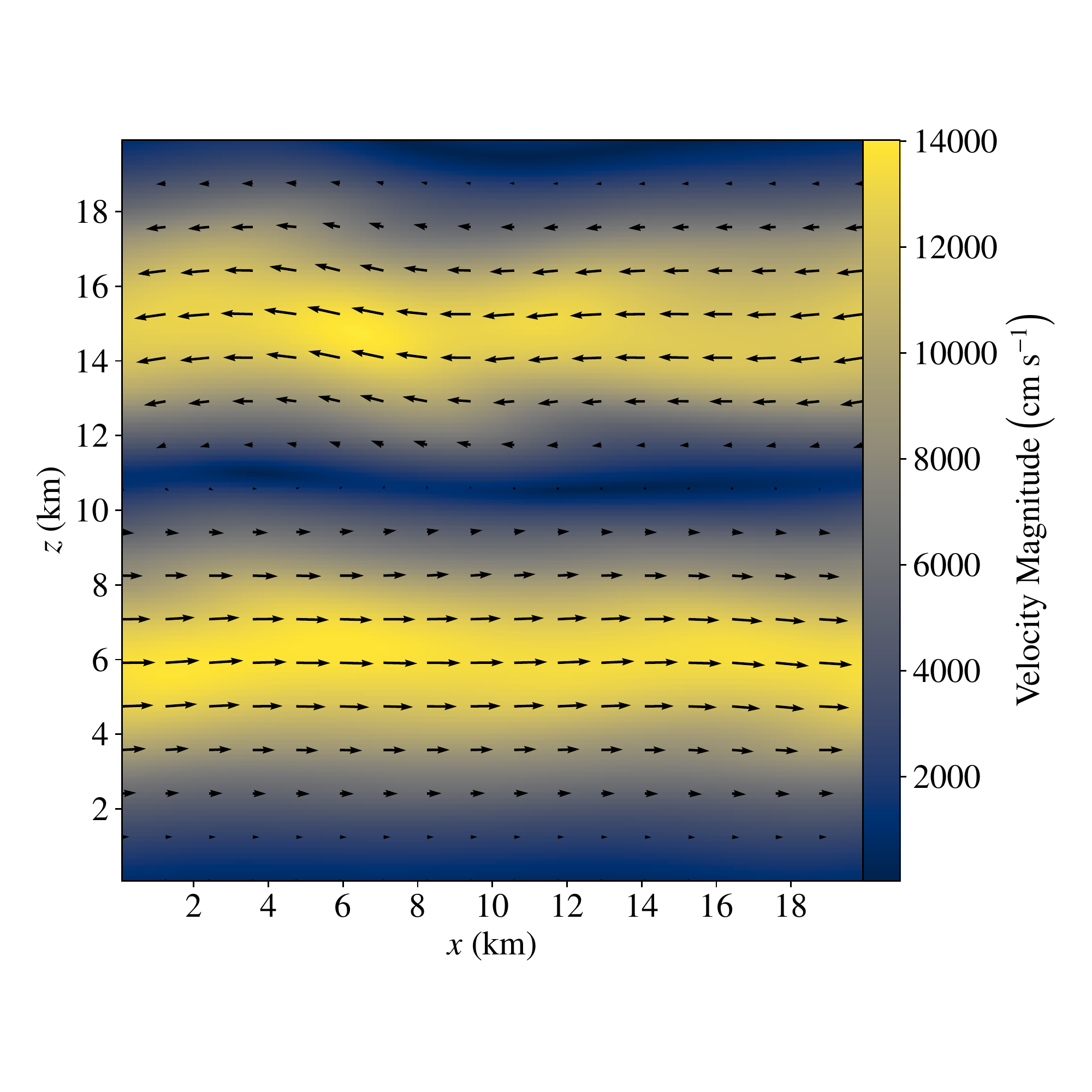}
 \includegraphics[width=0.45\textwidth]{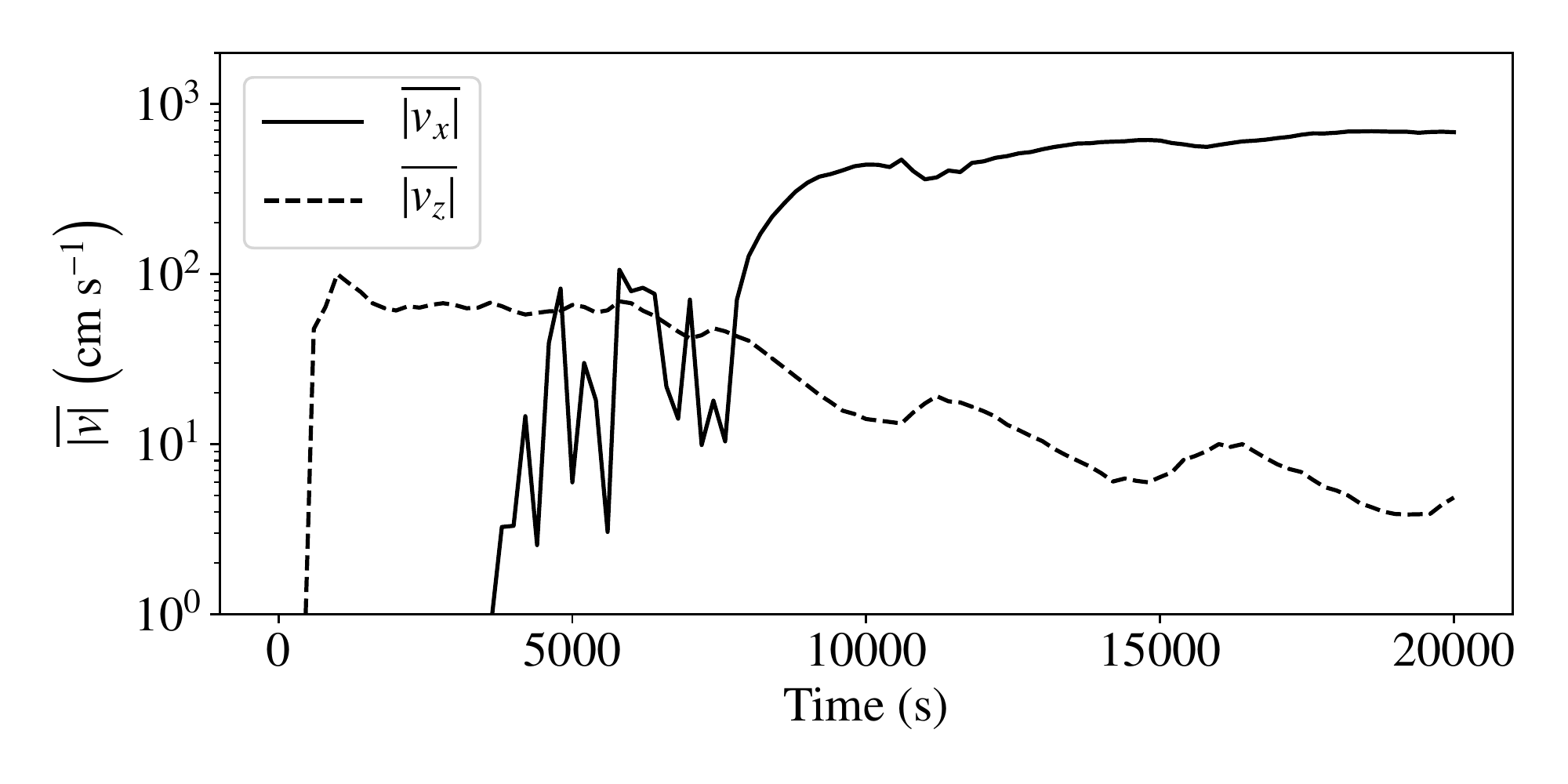}
 \caption{Illustration of the development of shear in 2D simulaitons. Top: Velocity map from a 2D simulation illustrating the transition to shear. The quivers indicate the velocity direction and magnitude. There are two distinct regions of oppositely directed flow separated by a transition of approximately equal width. Bottom: Time series of $\overline{|v_{x}|}$ and $\overline{|v_{z}|}$ showing the decay of the vertical component and the growth and saturation of the horizontal component. After a sufficient length of time (twice that of the 3D simulations), steady state has been reached and any convective motion is suppressed in favour of shear. \label{fig:2D_shear}}
\end{figure}

\begin{figure*}
 \centering
 \includegraphics[width=\textwidth]{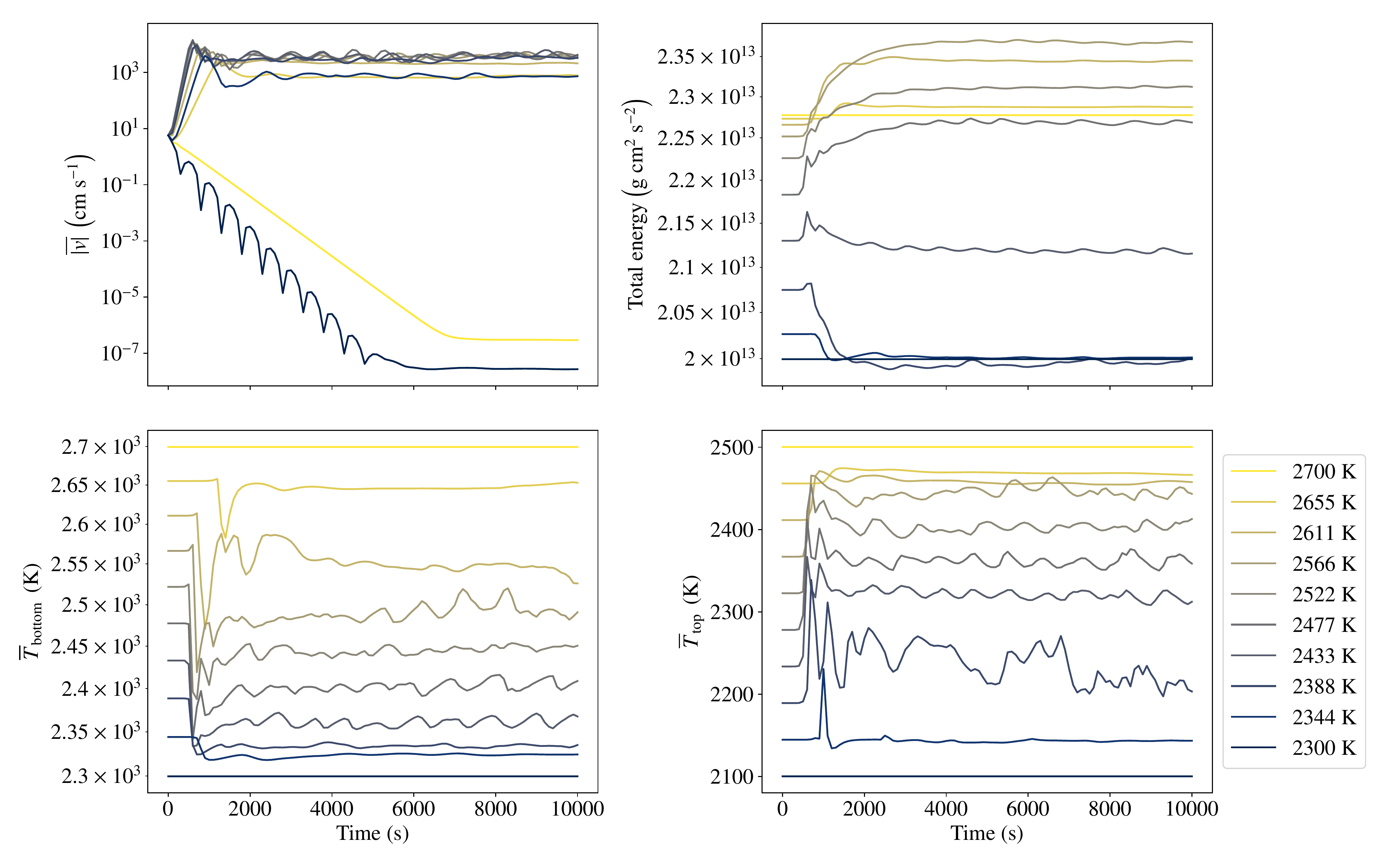}
 \caption{Time evolution of bulk fluid quantities over the course of the parameter study simulations. For each of the simulated deep-atmosphere temperatures (coloured from yellow to deep blue according to decreasing temperature), the average velocity magnitude, total energy, average deep-atmosphere temperature, and average temperature of the upper atmosphere are plotted. The behaviour of the average velocity magnitude is the primary metric for determining whether steady state has been reached. This occurs at 7000 s. \label{fig:time_evolution}}
\end{figure*}

\subsubsection{Numerical algorithm and grid}

The well-balanced all-regime method employed by ARK is a conservative, first-order accurate split scheme. This splitting allows the transport and acoustic terms to be solved separately with the gravitational acceleration added to the acoustic part, allowing for machine-accurate hydrostatic balance; full details of the method and its implementation in ARK are available in \cite{Chalons2016a} and \cite{Padioleau2019}.

The study was executed in two distinct phases. First, a parameter study in which a deep-atmosphere temperature range of $2300 - 2700 \rm{K}$ was covered, with ten separate temperature steps. This increment number was chosen to be a balance between computation cost and parameter coverage necessary to illustrate the diabatic convection's dependence on the deep-atmosphere temperature. The numerical grid used for this parameter study was $256^{3}$ and was kept constant for all values of deep-atmosphere temperature.

In the second phase, convergence was tested. This was achieved in separate resolution jumps where the resolution was doubled each jump from an initial low resolution of 320$^{3}$. This was done to relax the simulation into higher resolutions without introducing blocking effects, where one cell is replaced by more than eight new cells at a time. The resolution jumps are as follows: 310$^{3}$, 620$^{3}$, 1240$^{3}$, 2480$^{3}$, and 4960$^{3}$.

Due to the computationally intensive nature of the study, it is worth noting several steps that were taken to overcome the unique challenges encountered when performing simulations at this scale. At full resolution, the simulation grid was 5.2 TB in size (4960$^{3}$ grid cells, each containing 6 state variables at double precision accuracy). As such, the number of snapshots possible was restricted and, therefore, time series analysis required us to implement routines for the on-the-fly analysis of global variables. In addition, for visual inspection, a slicing routine was added that outputs the $x$-$y$, $x$-$z$, and $y$-$z$ 2D planes only, significantly reducing storage overhead and enabling high frame rate videos of the simulation to be produced.

Next we present the simulation results beginning with a stability analysis followed by the time series analysis with a discussion on the observable implication of the results, finally the resolution study is presented.

\section{Results and discussion}\label{sect:result}

First we determine the system's susceptibility to convection (both Schwarzschild and Ledoux) in the initial and steady state conditions. We then analyse the time series results and how they impact possible observables relevant to future space missions. We conclude by examining the system's convective properties at high resolution.

\subsection{Convective stability}
\label{sec:stability}

Applying the stability criterion of Eq. (\ref{eq:ledoux}) and (\ref{eq:schwarzschild}) to the initial conditions and to the final steady state system, produces the results shown in Fig. \ref{fig:criterion}. At all temperatures and altitudes the initial conditions are stable to Schwarzschild convection. This is by design, as the system needs to be Schwarzschild stable to exclude this type of convection from the simulations such that our analysis can be done purely in the Ledoux regime. Lower temperatures and higher altitudes are displaying greater stability.

In contrast, the system is susceptible to Ledoux convection for a portion of the simulation domain for every temperature. The maximum altitude of the convectively susceptible region increases as the deep-atmosphere temperature increase. This dependence is due to the mass-mixing ratio being a direct function of the deep-atmosphere temperature and temperature range of the study. A pattern directly mirrored from the Schwarzschild criterion case above. However, in the case of the Ledoux criteria, the system exceeds zero at every deep-atmosphere temperate value. This observation makes clear the impact of the addition of the mean-molecular-weight gradient on the system's convective stability.

This global behaviour is expected due to the manner in which the mean molecular weight is prescribed, as a linearly increasing profile that transitions from low to high altitude as the imposed deep-atmosphere temperature is progressively increased with each subsequent simulation. As the mean-molecular-weight gradient is the defining difference between the two convective criteria (Eqs. (\ref{eq:ledoux}) and (\ref{eq:schwarzschild})) and as the system is Schwarzschild stable both initial and for all $t > 0$, we can conclude that any convection established is solely due to the imposed mean-molecular-weight gradient and maintained by the chemical and thermal source terms.

The same criterion is calculated for the system at $t=t_{\rm{max}}$ and is shown in the bottom row of Fig. \ref{fig:criterion}. We can see that the susceptibility to both Schwarzschild and Ledoux convection, specified in the initial conditions, is maintained through to the end of the simulation. This illustrates the lack of stabilising or destabilising influences on the system beyond those we impose through the initial conditions and the source Eqs. (\ref{eq:delta_t}) and (\ref{eq:delta_x}).

As an aside, when running prototype simulations in 2D, the system was found to readily transition to shear; under the same conditions where the 3D simulations do not. Figure \ref{fig:2D_shear} illustrates this 2D shear profile with two distinct higher velocity bands running perpendicular to the direction of gravity and opposite to each other. A narrow boundary layer exists between the two flow streams where the velocity falls by an order of magnitude. The horizontal and vertical velocity components are also plotted below the velocity map in Fig. \ref{fig:2D_shear} and highlight the horizontal velocity's growth and the degree to which shear has dominated the simulation at late times. {These 2D shear modes have already been observed in the context of double-diffusive convection \citep{Garaud2015} and seems to be a result of the dimensionality of the setup, an effect that we also identify in this paper.}

From a physical perspective, this system should be susceptible to mixing via the Kelvin-Helmholtz instability. However, from a numerical point of view, the simulation is overall first-order accurate. The Riemann solver employed was similar to the Harten-Lax-van Leer-Contact (HLLC) type, known to be necessary for capturing this type of instability; however, as the spatial reconstruction and time integration is first order, the degree of dissipation in the model may simply be too great for the instability to establish. This is, however, purely speculative and we leave further discussion of the role of shear to a future publication in which the general case of shear in 3D simulations will be studied \citep[see also][]{Garaud2015, Garaud2019}.

With the stability of the system and the convective regime established we will analyse the time-dependent nature of the simulations.

\begin{figure*}
 \centering
 \includegraphics[width=\textwidth]{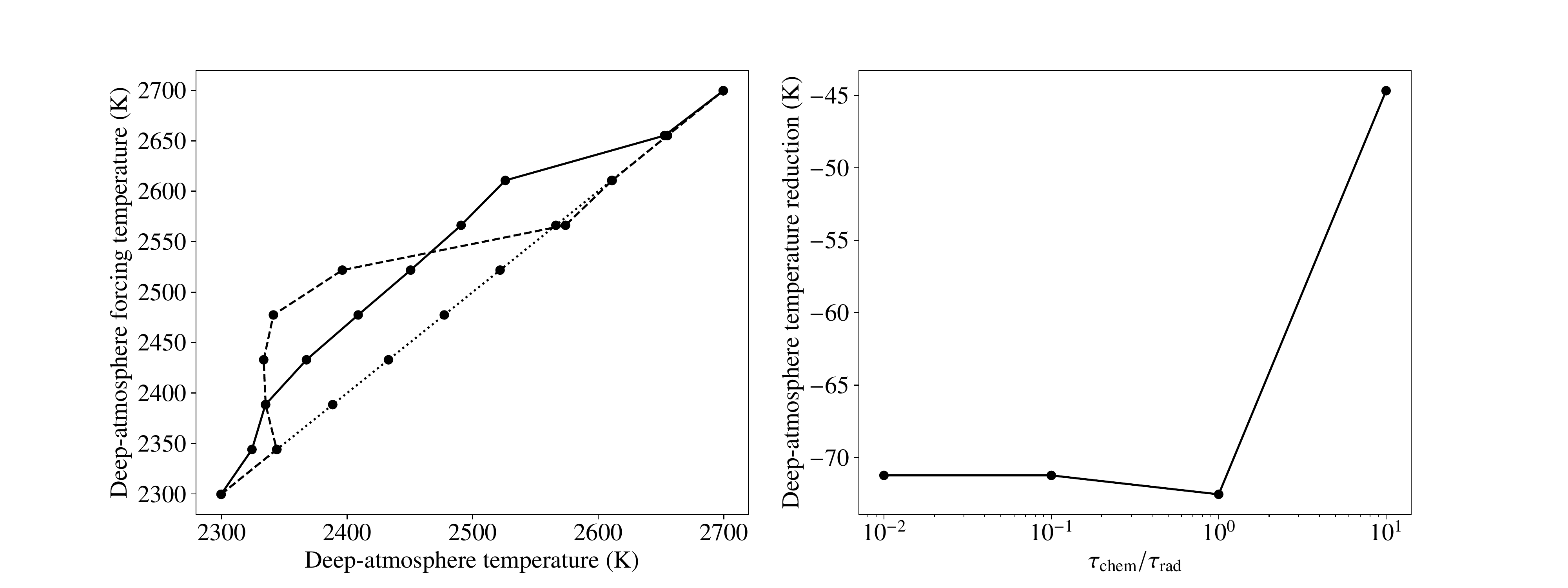}
 \caption{Deep-atmosphere forcing (equilibrium) temperature and the ratio of timescales explored, illustrating convergence. Left: Relationship between the deep-atmosphere forcing (equilibrium) temperature and the resulting deep-atmosphere temperature. The dotted curve represents the initial adiabatic conditions. The solid curve illustrates the deviation from this due to the convective motion. The final profile shows a departure from the adiabatic branch towards the diabatic branch. The dashed curve highlights a transient early stage when the perturbation towards the diabatic branch is most pronounced. Right: Ratio of the forcing timescales $\tau_{\rm{chem}}/\tau_{\rm{rad}}$ for $T_{0}~=~2522 \ \rm{K}$ with $\tau_{\rm{chem}}~=~1, 10, 100, \ \rm{and} \ 1000 \ \rm{s}$; $\tau_{\rm{rad}}~=~100 \ \rm{s}$ for all values. Convergence is reached once $\tau_{\rm{chem}}/\tau_{\rm{rad}}~<~1$. \label{fig:FT_curve}}
\end{figure*}

\subsection{Time series}

A time series of the parameter study is shown in Fig. \ref{fig:time_evolution}. To avoid confusion, here it is worth reiterating the difference between what is meant by average bottom temperature, $\overline{T}_{\rm{bottom}}$, a time-dependent quantity measured from the simulations; and the forced deep-atmosphere temperature, $T_{0}$, which is a parameter. Each separate coloured line in the legend of Fig. \ref{fig:time_evolution} refers to a different forced deep temperature, $T_{0}$.

The velocity magnitude, top left, was used as a steady state measure. Once this quantity becomes approximately constant with time, we assume steady state has been achieved. As the initial conditions are seeded by a convective perturbation, if the system itself is unstable to convection (see Sect. \ref{sec:stability}), then this initial perturbation will grow rapidly until saturated at a maximum value. Once reached, the velocity magnitude will slightly oscillate about this maximum. However, if the system is stable to convection, then the initial perturbation will decay away steadily until swamped by numerical noise.

Both of the above behaviours are seen in Fig. \ref{fig:time_evolution}, with both extremes in the deep-atmosphere temperature (2300 K and 2700 K) displaying a decaying velocity magnitude. The velocity magnitude grew in every other simulation until saturation, which occurred at \textasciitilde1000 s for all convectively unstable simulations.

The right plot of the top row of Fig. \ref{fig:time_evolution} shows the total energy evolution. Departure from constant energy may at first seem unexpected; however, it is physically justified as the simulation includes a source term and boundary conditions are not reciprocal on all faces. Indeed the $z$-direction boundary conditions are customised to produce the simulation conditions we seek. If all the boundary conditions were reciprocal and source terms set to zero, then the total energy would remain constant as the numerical scheme is conservative.

The top atmosphere temperature, $\overline{T}_{\rm{top}}$, exhibits an increase across the $T_{0}$ range (except in the extremes 2300 K and 2700 K). This increase, together with the decrease in $\overline{T}_{\rm{bottom}}$, means a decrease in the vertical temperature gradient. This result indicates that the presence of convection directly leads to a reduction in the upper atmosphere temperature with respect to that specified initially and that which would be maintained through the forcing profile.

As the temperatures $T_{0} = 2300 \rm{K}$ and $T_{0} = 2700 \rm{K}$ do not experience this temperature-gradient reduction -- instead maintaining the hydrostatic equilibrium imposed initially but nonetheless allowed to evolve dynamically -- we can conclude that convection in this model atmosphere is responsible for modifying the temperature structure. In the case of bulk planetary atmospheres, this will result in an observable upper-atmosphere temperature corresponding to a deep-atmosphere temperature (if hydrostatic equilibrium is assumed), which is larger than that actually present. This notion will be explored in greater depth in Sect. \ref{sec:obs}.

The division between the convectively stable (2300 K and 2700 K) and unstable (2344 K $\rightarrow$ 2655 K) $T_{0}$ simulations is consistent in all the bulk quantities tracked throughout the evolution. The mid temperatures show the greatest degree of energy increase into the system. This is evident through the largest temperature changes at the top and bottom of the simulation domain.

\subsection{Thermal forcing versus deep-atmosphere temperature}
\label{sec:FT}

In the absence of convection, the temperature profile of the model would be completely imposed by the Newtonian forcing profile, for example in the radiative part of the atmosphere. The deep-atmosphere temperature would be simply given by $T_{0}(z_\mathrm{bottom})$. If we plot this forcing deep-atmosphere temperature against the final deep-atmosphere temperature, we can analyse the impact that convection has on the temperature profile. This is displayed in the left hand plot of Fig. \ref{fig:FT_curve} for initial ($t=0$ s), intermediate ($t=500$ s) and steady state ($t=10000$ s) time states.

If we assume that the region we are probing is stable but close to the radiative-convective boundary in the planet, we can assume that the energy flux is well approximated by $\phi_{\rm{base}} = \sigma_{\rm{SB}} T_{0}(z_\mathrm{bottom})^{4}$. As a consequence, this energy flux is also a measure of the equivalent forcing convective flux in the deep atmosphere in a model following radiative-convective equilibrium. We are not explicitly forcing an energy flux at the bottom of the domain; however, the deep-atmosphere forcing temperature can, to some extend, be seen as a proxy for a forcing energy flux in our setup.

The dotted (initial condition adiabatic profile) and solid curves illustrate the difference in behaviour for the adiabatic and diabatic convecting systems, respectively. The system effectively undergoes a bifurcation, as the convective motion builds and the instability saturates. The deep-atmosphere temperature is forced away from the Newtonian (sub-adiabatic) profile towards the diabatic profile in the mid temperatures, where the mean-molecular-weight gradient is greatest. This effect is most apparent early in the simulation while the instability is still growing and before saturation. The dashed curve ($t=500$ s) in Fig. \ref{fig:FT_curve} shows this initial phase and has a much more pronounced bifurcation than the steady state solution.

{The behaviour described above means that the net effect of the convective instability (and therefore the presence of a mean-molecular-weight gradient), is that the deep-atmosphere temperature is cooler than the setup without mean-molecular-weight gradient. It is clear then that the upper atmosphere possesses a greater temperature than that allowed by thermal forcing and the deep-atmosphere temperature.}

The bifurcation in the temperature plot in Fig. \ref{fig:FT_curve} is similar to the bifurcation proposed in \cite{Tremblin2019} in the case of the O-CH$_4$ transition in the atmosphere of brown dwarfs and extrasolar giant exoplanets. It therefore suggests that the analogy to the Nukiyama curve for the boiling crisis experiment \citep{Nukiyama1934} is relevant and can be used to deduce bifurcation in the behaviour of convective systems in general.

The ratio of the forcing timescales, $\tau_{\rm{chem}}/\tau_{\rm{rad}}$, determines the convergence of the deep-atmosphere temperature perturbation away from the adiabatic profile. To test this convergence, multiple iterations of the $T_{0}~=~2522 \rm{K}$ simulation were conducted with $\tau_{\rm{chem}}~=~1, 10, 100 \ \rm{and} \ 1000 \ \rm{s}$. The right plot of Fig. \ref{fig:FT_curve} shows the deep-atmosphere temperature as a function of $\tau_{\rm{chem}}$; in the interest of brevity $\tau_{\rm{rad}}~=~100 \ \rm{s}$ for all the simulations and only $\tau_{\rm{chem}}$ was varied. As can be seen, the deep-atmosphere temperature does not undergo any further reduction once $\tau_{\rm{chem}}/\tau_{\rm{rad}}~<~1$, and converges at $\sim 70 \ \rm{K}$ less than the initial condition temperature of $2522 \ \rm{K}$. {It also shows that the reduced temperature gradient is not present if the chemical timescale is too slow.}

\subsection{Convergence}

\begin{figure*}
 \centering
 \includegraphics[width=\textwidth]{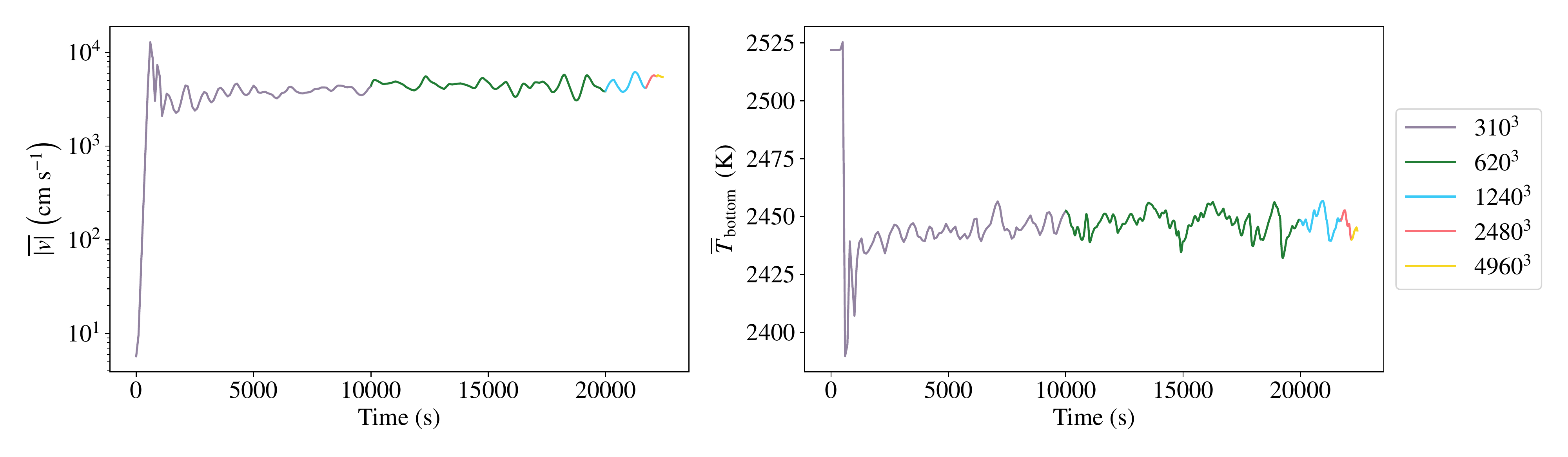}
 \caption{Time evolution of bulk fluid quantities for the resolution study. One can see that each successive resolution jump leaves the bulk velocity magnitude and deep-atmosphere temperature unchanged. This demonstrates  that convergence is achieved at the low resolutions used in the parameter study. \label{fig:time_evolution_GC}}

\end{figure*}

\begin{figure}
 \centering
 \includegraphics[width=0.5\textwidth]{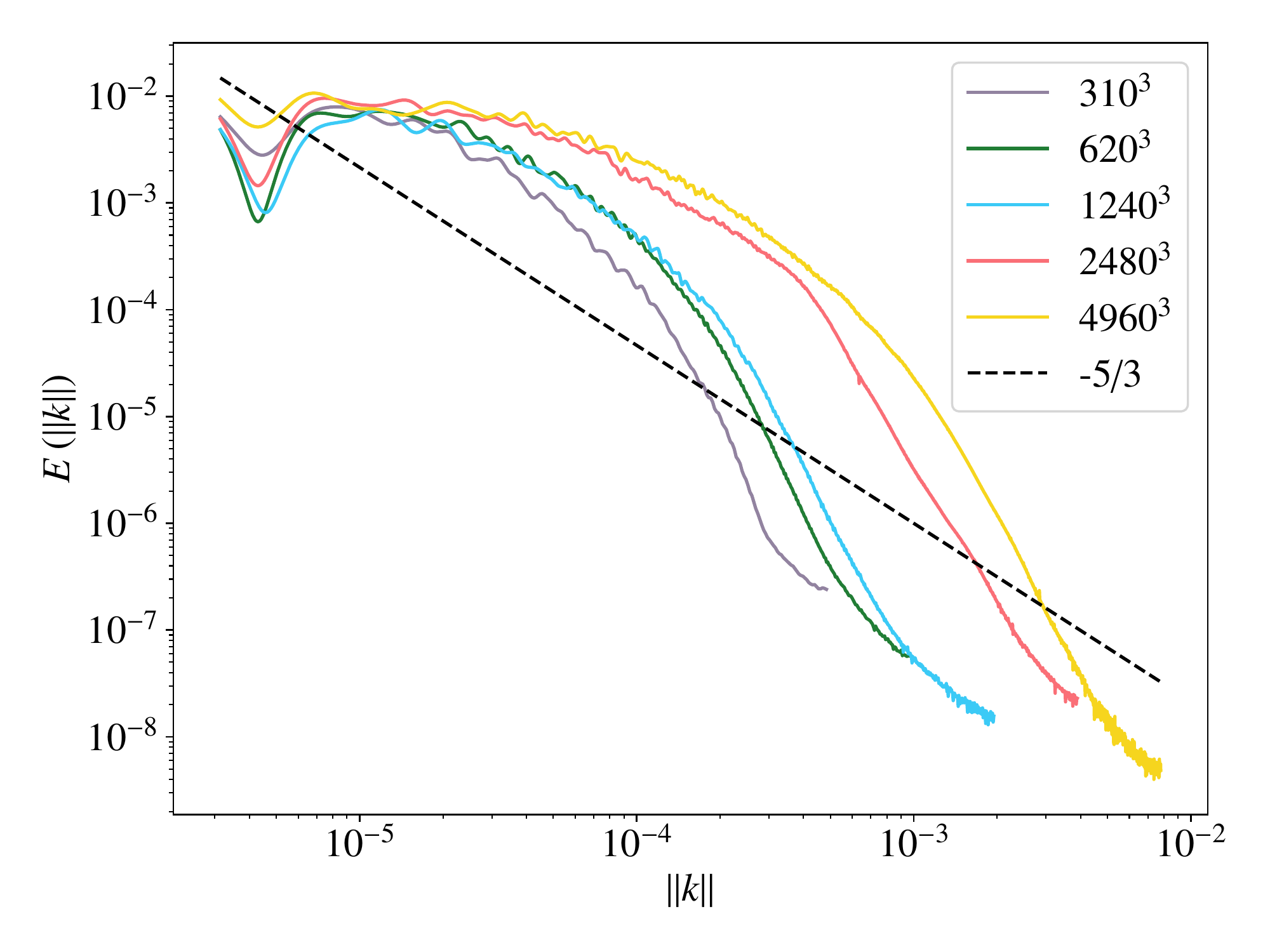}
 \caption{Kinetic energy power spectrum for each resolution jump used in the convergence study. The Kolmogorov spectrum, $E(k)~\propto~k^{-5/3}$, is shown as the dashed black line for illustrative purposes. The $E(||k||)$ is calculated in 2D at the midplane in the $z$ direction. All resolutions converge at small $||k||$ (large length scales), indicating that all resolutions capture the large-scale convective behaviour. At large $||k||$ (small length scales), we see the cascading of kinetic energy characteristic of turbulence. At this point the different simulations begin to diverge as the additional resolution resolves the smaller scales to differing degrees in the simulations. \label{fig:power_spec}}

\end{figure}

\begin{figure*}
 \centering
 \includegraphics[width=\textwidth]{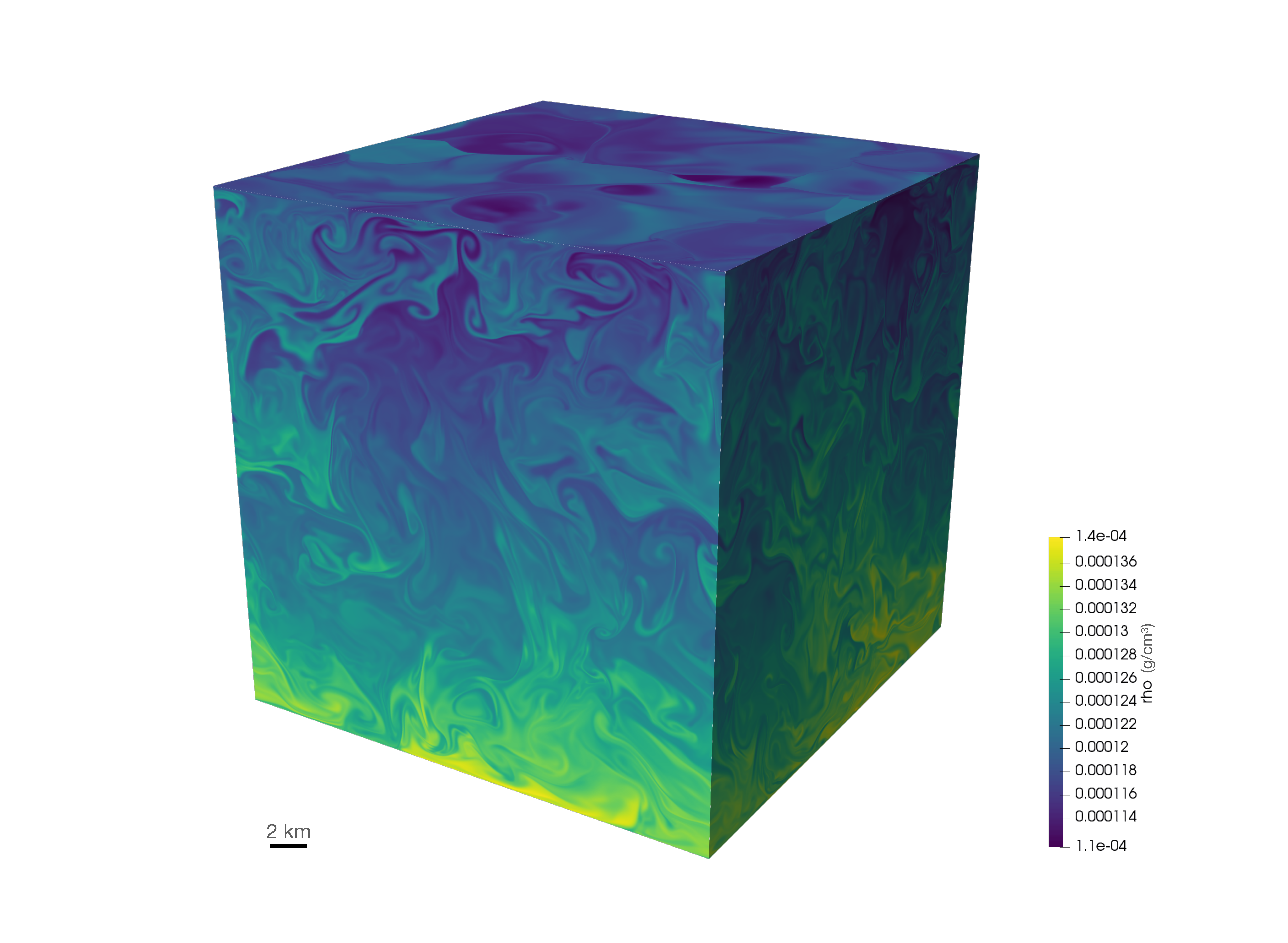}
 \caption{Rendering of the density of the high resolution simulation's entire computational volume. Units are in cgs. Both large- and small-scale features are evident, including Rayleigh-Taylor-like and Kelvin-Helmholtz-like instabilities. Despite the high degree of mixing via these turbulent processes, the large-scale vertical density gradient is still apparent and the convective motion is persistent. \label{fig:volume}}
\end{figure*}

To test the convergent properties of our computational setup, we selected a deep-atmosphere temperature in the middle of the temperature range, $T_{0} = 2522$ K. We performed a set of simulations, varying the resolution between $310^{3} \rightarrow 4960^{3}$. The lowest resolution is different from the resolution of the time series study simply to provide a value that can be recursively doubled-up to a sufficiently high resolution while simultaneously allowing for parallel decomposition of the computational domain. A snapshot of the final resolution ($4960^{3}$) is displayed in Fig. \ref{fig:volume}, illustrating the turbulent fluid present in the simulation beyond that of the bulk convective motion.

To asses the impact of the resolution jumps and to ensure convergence at each intermediate stage, the average velocity magnitude and deep-atmosphere temperature were tracked. The results are displayed in Fig. \ref{fig:time_evolution_GC} with each colour indicative of a different resolution. It can clearly be seen that the resolution jumps have no impact on these bulk quantities. With the velocity magnitude maintaining the convectively unstable profile. Indeed every resolution leaves the maximum velocity magnitude unchanged. This narrative persists in the case of the deep-atmosphere temperature, where the reduction due to the action of convection remains consistent across every resolution jump. It is also worth noting that the amplitudes of the fluctuations are around $\sim$20~K, that is, approximately 25\% of the overall temperature reduction. These fluctuations in the case of CO-CH$_4$ radiative convection \citep{Tremblin2020} could potentially explain rotational spectral modulations in the atmosphere of brown dwarfs \citep{Buenzli2012,Apai2013,Biller2017,Artigau2018}

Finally, to illustrate this separation of length and energy scales in relation to the convective instability and turbulent higher order behaviour, we have plotted in Fig. \ref{fig:power_spec} the kinetic energy power spectrum, $E(||k||)$. These data are taken from the $x$-$y$ plane at the midpoint in the $z$ direction, for each resolution in the convergence study. Overlaid is also the Kolmogorov spectrum, $E(k)~\propto~k^{-5/3}$, for comparison.

At large length scales (small $||k||$), each resolution results in approximately the same contribution to $E(||k||)$; however, there is an anomalous dip for all resolutions were $10^{-5}~<~||k||$ ($x~\lesssim~6.3 \rm{km}$), At these scales, the bounding box of the simulation starts to be probed. As the reciprocal boundaries used in the simulation are not reflected in the calculation of $E(||k||)$, when the smallest vales of $||k||$ are probed there is only contributions from the box corners and hence an incomplete picture is obtained.

For $10^{-5}~<~||k||~<~10^{-4}$ ($0.63~\rm{km}~\lesssim~x~\lesssim~6.3~\rm{km}$), we initially see the same $E(||k||)$ for every resolution. This result is expected as the largest cell separation is $\sim 0.065$ km (resolution of $310^{3}$), meaning that wave numbers $||k||~\approx~10^{-5}$ are spanned by $\sim10$ numerical cells.

Only when $10^{-4} < ||k||$ ($x \lesssim 0.63$ km) do we see the impact of higher resolution and the additional turbulent modes that this allows for. The viscous dissipation at the lower limit is numerical in nature, this means that as kinetic energy cascades from large to small eddies, the cell width sets the lower limit on the length scale of the turbulent eddies.

Temperature stratification of the atmosphere is the focus of this study and as this is impacted by the large-scale convective motion, it is tempting to dismiss the role of small-scale structures. However, turbulence introduces instabilities that have the potential to disrupt large-scale behaviour, for example the transition to shear modes \citep[see e.g.][]{Garaud2015}. This does not occur, and the unlocking of this kinetic energy cascade with higher resolution does not result in disruption of the convective behaviour. This indicates that small-scale dynamics do not inhibit bulk convective motion and inhabit separate regions of the kinetic energy power spectrum.

\subsection{Observational implications}
\label{sec:obs}

\begin{figure*}
 \centering
 \includegraphics[width=0.45\textwidth]{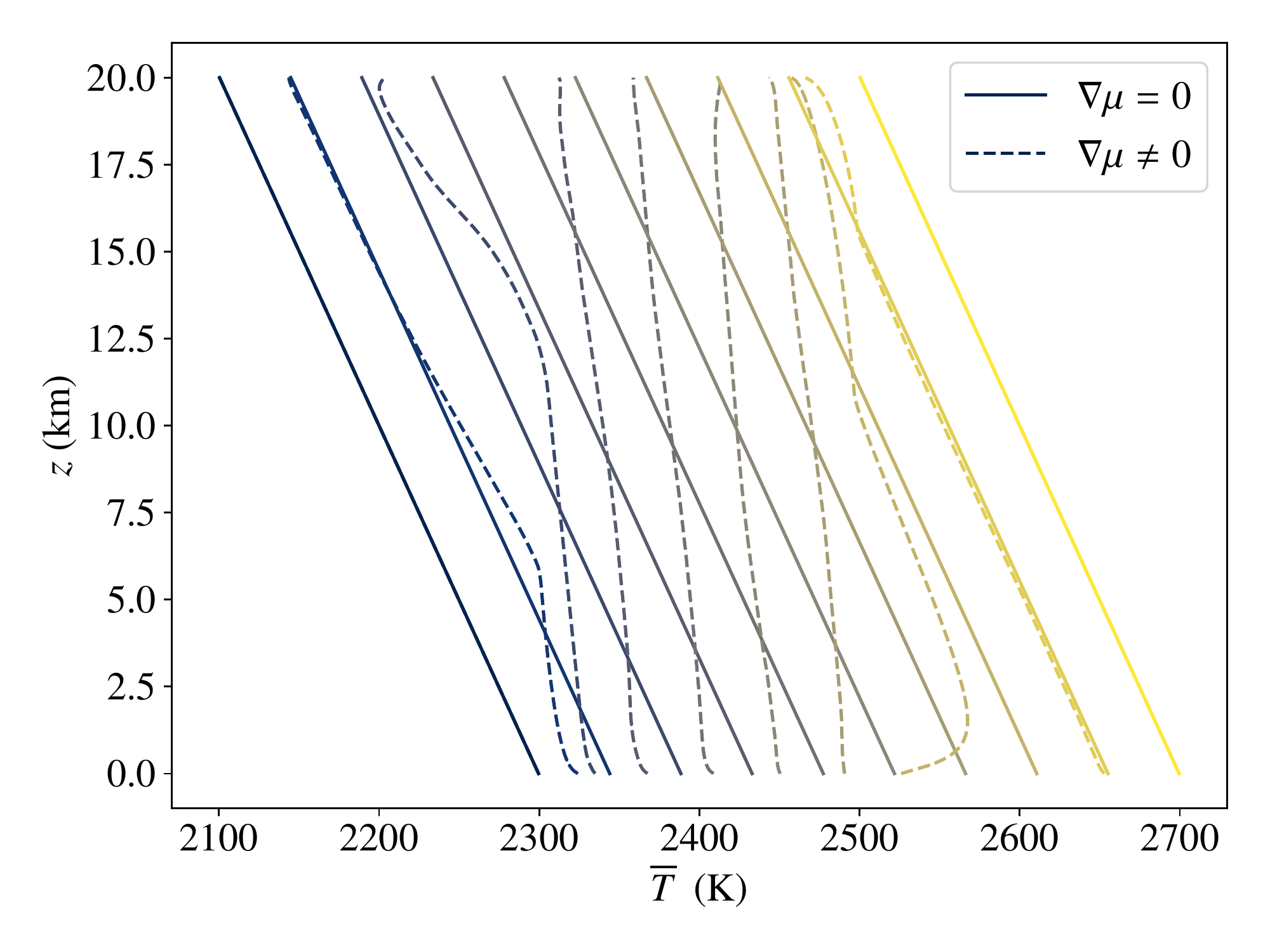}
 \includegraphics[width=0.45\textwidth]{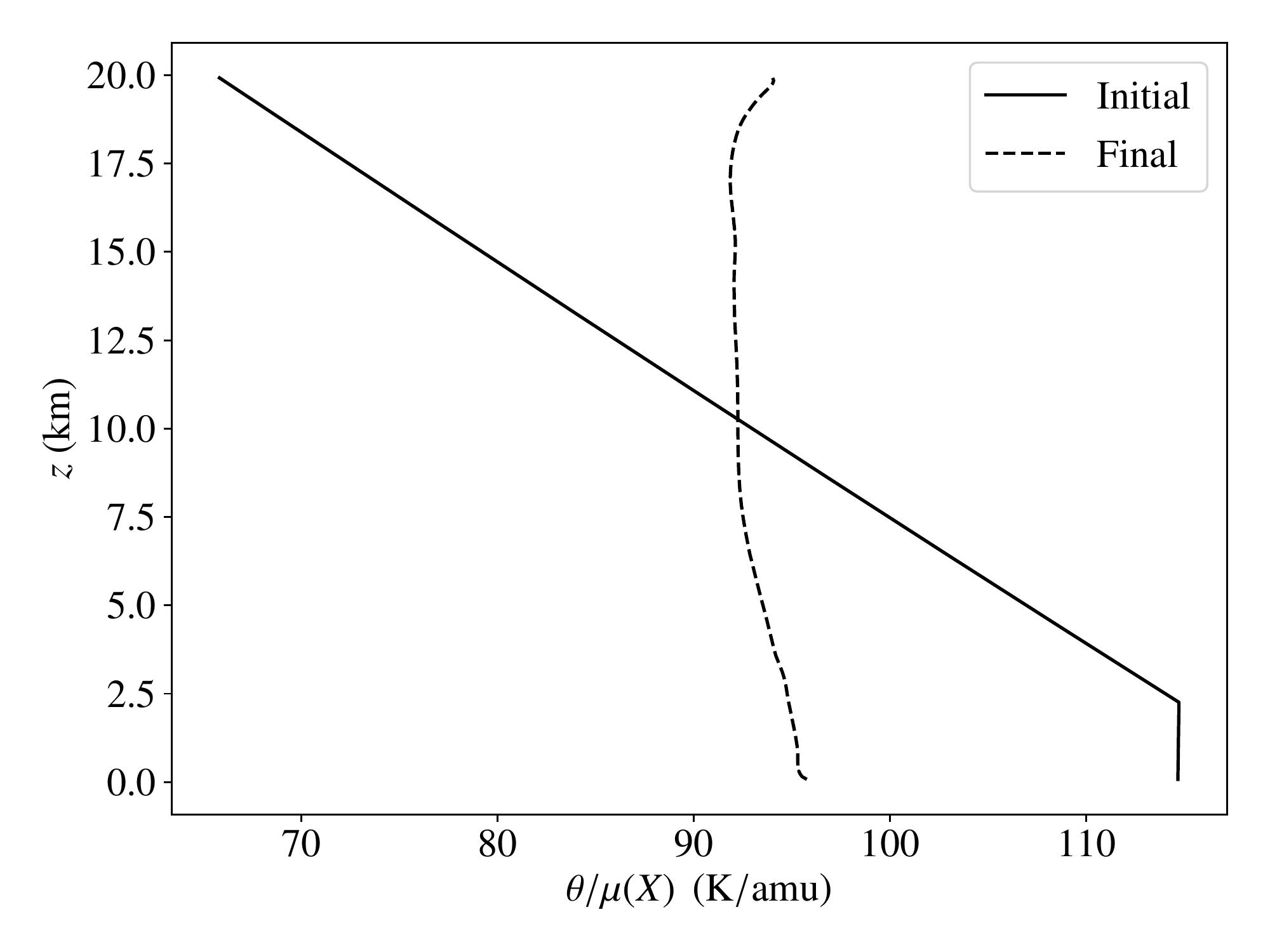}
 \caption{Temperature and potential temperature profiles illustrating the deviation from the adiabatic behaviour. Left: Vertical temperature-gradient profiles comparing the adiabatic (solid line) to the diabatic (dashed line), corresponding to the dotted and solid curves, respectively, in the left hand plot. The colouring indicates a separate simulation, each with a different initial deep-atmosphere temperature. Each of these profiles correspond to a data point on the curves in the left hand plot. The central temperatures show the largest deviation from the adiabatic profile. Right: Vertical profiles of potential temperature for both the initial and final conditions of the $T_{0}=2522 \ \rm{K}$ simulation. The impact of the mean-molecular-weight gradient is to homogenise entropy
 \label{fig:temp_comp}}
\end{figure*}

In order to place the above results into the context of observational interpretation we {compare the expectations from thermal forcing alone to the simulated curve, whose data comes directly from the simulation domain and assumes nothing about the thermal profile. we can see there is considerable difference over the 20 km of simulated atmosphere with the most pronounced difference being $\sim 160$ K for $T_{0}~\approx~2400$ K in Fig.~\ref{fig:FT_curve}.{This is illustrated more concisely in the left-hand plot of Fig. \ref{fig:temp_comp} where we can see the temperature gradient is modified by the presence of a mean-molecular-weight gradient. This modification moves the system towards a constant entropy profile. This can be seen in the right hand-side of Fig. \ref{fig:temp_comp}, were we plot the profile of the ratio of potential temperature to mean molecular weight,
\begin{equation}
    \frac{\theta(z)}{\mu(X)} = \frac{T(z)}{{\mu(X)}} \left( \frac{P_{0}}{P(z)} \right)^{\frac{\gamma - 1}{\gamma}}.
\end{equation}
 The point is that the reduced temperature gradient occurs even in the situation where the entropy is homogenised (a point argued in \cite{Tremblin2019}).}

For the parameters used in this study, the scale height is $H_{p} \sim 70$ km. Therefore, over approximately a third of the scale height, the temperature deviation from the adiabatic profile is up to $\sim 6 \%$ in the most pronounce case. It should be noted though that this effect is only present in the portion of the atmosphere where there is sufficient mean-molecular-weight gradient to promote convective motion. This $\sim 6\%$ temperature deviation will only impact this portion of the atmosphere depending on the chemical timescale used in the simulation, the overall temperature reduction across other sections of the atmosphere, at different altitudes, will depend on this chemical timescale and could be higher or lower.

The implication for observations of planetary atmospheres, is that the chemical composition and therefore the potential for molecular weight gradients need to be considered before assumptions can be made about the deep-atmosphere and overall temperature profiles. Convective instabilities such as Ledoux convection, driven by chemical and thermal source terms, as investigated here, result directly from these composition gradients.

{Turning the above analysis upside down, the temperature of the upper atmosphere is often placed in the context of that which would be produced by the thermal transfer of energy from the deep  to the upper atmosphere. Therefore, the presence of source terms and mean-molecular-weight gradients allows the upper atmosphere to be considerably hotter than that what would be afforded by a deep-atmosphere temperature according to the thermal forcing alone.}

This potential for a {reduced temperature gradient} directly mirrors the so called reddening observed in the spectra of brown dwarfs \citep[traditionally explained with silicate clouds]{Tsuji1996,Chabrier2000,Allard2001,Marley2010}, where the deep-atmosphere infrared flux is found to be smaller than expected based on adiabatically extrapolating the observed upper-atmosphere temperature down to the deep atmosphere. It has been proposed by \cite{Tremblin2016,Tremblin2019} that these theories overestimate the deep-atmosphere temperature by assuming an adiabatic convective profile. Diabatic convection provides therefore a compelling mechanism for accounting for this reddening in brown dwarfs. This type of behaviour could be at the origin of the need for a reduced temperature gradient in the atmosphere of cold brown dwarfs of spectral types T and Y \citep[induced there by the transition between N$_2$ and NH$_3$,][]{Leggett2017,Leggett2019}, with growing evidence that the increased mid-infrared flux cannot be explained by clouds for these objects. {However, we emphasise again that the temperature-gradient reduction in the Ledoux regime is of a different nature since it happens even if the simulation is nearly adiabatic at saturation. The common point is that we can obtain reduced temperature gradients even when the temperature gradient is stabilising and this effect can have observational implications.}

We have seen from the above simulations that the chemical transition of CO+O~$\leftrightarrow$~CO$_{2}$, and their separation into atmospheric layers with the associated boundary between them, is capable of providing the mean-molecular-weight gradient required for the establishment of Ledoux convection. Both CO$_{2}$ and CO are readily available in the atmospheres of terrestrial exoplanets, especially in the early stages of atmospheric formation. We therefore {speculate} that the mechanism responsible for the reddening observed in the spectra of brown dwarfs {can} also be present in the spectra of terrestrial exoplanetary atmospheres in the form of reduced temperature gradients in the Ledoux regime. This type of mechanism can also be activated in the presence of other chemical transitions, or other chemical and thermal source terms in general, following the general theory developed in \cite{Tremblin2019} {and should not be excluded a priori. The setup used in this paper is, however, idealised: We list in the next subsection the limitations of this work for more realistic applications.}

The spectra and temperature of exoplanetary atmospheres will feature in the future observational objectives of instruments such as JWST and the European Extremely Large Telescope. It is therefore vital that the results are interpreted with robust theoretical models. We therefore make the case that the role of chemically and thermally driven convection needs to be incorporated into any analysis of observations concerned with planetary atmospheres.

\subsection{Limitations}

We recap here the approximations and the three limitations of the idealised setup used in this paper. First, the temperature and compositional gradients need to be evaluated from detailed 1D models with all the physics (chemistry, opacities) relevant for rocky-exoplanet secondary atmospheres. Second, the chemical timescale needs to be evaluated with chemical networks validated for the conditions of rocky-exoplanet secondary atmospheres. Third, the EOS is oversimplified here, keeping a constant adiabatic index in order to finely control the convective behaviour: A more realistic EOS needs to be used.

For the last point, we highlight that this is not only a question of varying the adiabatic index with composition: since we are in a reactive fluid, the heat capacity has a reactive contribution well-defined when the system is close to chemical equilibrium. This leads to the following equation for $c_v$:
\begin{eqnarray}
c_v &=&\left(\frac{\partial e}{\partial T}\right)_V\cr
&=& X_\mathrm{eq} c_{v_1} +(1-X_\mathrm{eq}) c_{v_2} + \frac{d X_\mathrm{eq}(T)}{dT} (e_1-e_2)
.\end{eqnarray}
The resulting adiabatic index can then be deduced from the relation $\gamma-1 = k_\mathrm{b}/(c_v \mu)$. This form should be relatively valid in our setup since we use a fast chemical timescale; however, we point out that these thermodynamic quantities are likely challenging to define properly when the system is out of chemical equilibrium \citep[see e.g.][]{Lebon2008}.

The setup used in this paper is therefore relatively idealised. It, however, shows that we cannot exclude a priori the occurrence of a reduced temperature gradient in these atmospheres.

\section{Conclusions}

We have numerically confirmed the ability of compositional and thermal source terms to induce a reduction in an atmospheric temperature gradient {in the Ledoux regime that could happen in the context of rocky exoplanet atmospheres.} This has been done via both time series analysis and a convergence study that spanned an order of magnitude in resolution. We have also explored the impact of 2D versus 3D in these setups, showing that the shear modes emerging in 2D do not impact the 3D simulations.

We have studied an idealised test case based on a  CO+O~$\leftrightarrow$~CO$_{2}$ transition, which has been used as a prototype chemical transition that is readily present in the atmospheres of hot rocky exoplanets. This makes it relevant not only for the study of present-day young or irradiated terrestrial exoplanet atmospheres but also paleo-climatology, as the thermal profiles and parameters used in this study are comparable to the early Earth's climate. We have used this transition to demonstrate that a mean-molecular-weight gradient reduces the atmosphere's temperature gradient, a result that does not converge away with resolution. We find that the above transition leads to a bifurcation of the atmosphere's temperature profile,  with the largest deviation coinciding with a maximum in the mean-molecular-weight gradient, a result that directly impacts how the deep-atmosphere temperature should be deduced from that of the upper atmosphere.

Although more detailed analyses with 1D models are needed, our results indicate the possibility of an analogy of the reddening (currently observed in the spectra of brown dwarfs) in the spectra of terrestrial exoplanet atmospheres. This is a new prediction for rocky exoplanet atmospheres and strongly motivates the use of diabatic models when interpreting atmospheric spectra. These types of observations will feature in future campaigns of instruments designed for the characterisation of the atmospheres of exoplanets, such as JWST and the European Extremely Large Telescope, making robust theoretical models vital for accurately interpreting observations.

\section{Acknowledgments}

We thank the referee for their helpful comments. S.D.Y. and P.T. acknowledge support by the European Research Council under grant agreement ATMO 757858. This work was granted access to the HPC resources of the supercomputer Joliot-Curie@TGCC under the allocation 2019-100925 made by GENCI and of the supercomputer Jean-Zay@IDRIS under the ``grand challenge'' 2019-100951.

\bibliographystyle{aa}
\bibliography{bibliography}

\end{document}